\RequirePackage{fix-cm}
\documentclass[smallextended]{svjour3}       
\smartqed  
\usepackage{graphicx}
\usepackage[colorlinks=true]{hyperref}
\usepackage[all]{hypcap} 
\usepackage{hyperref}
\def\sgra{Sgr~A$^*$}

\begin{document}

\title{Testing General Relativity with the Event Horizon Telescope}


\author{Dimitrios Psaltis}


\institute{D. Psaltis \at
  Department of Astronomy\\
  The University of Arizona\\
  933 N Cherry Ave, Tucson, AZ 85711, USA\\
              \email{dpsaltis@email.arizona.edu}           
}

\date{Received: date / Accepted: date}

\maketitle

\begin{abstract}

The Event Horizon Telescope is a millimeter VLBI array that aims to
take the first pictures of the black holes in the center of the Milky
Way and of the M87 galaxy, with horizon scale resolution. Measurements
of the shape and size of the shadows cast by the black holes on the
surrounding emission can test the cosmic censorship conjecture and
the no-hair theorem and may find evidence for classical effects of the
quantum structure of black holes. Observations of coherent structures
in the accretion flows may lead to accurate measurements of the spins
of the black holes and of other properties of their spacetimes. For
\sgra, the black hole in the center of the Milky Way, measurements of
the precession of stellar orbits and timing monitoring of orbiting
pulsars offer complementary avenues to the gravitational tests with
the Event Horizon Telescope.
  
\end{abstract}

\tableofcontents

\section{Introduction}

The very small size of a black hole makes direct imaging of its
environment with horizon-scale resolution a very challenging goal.
Taking a picture of the black hole in the center of the Milky Way,
which subtends the largest angle in the sky among all known black
holes, requires an angular resolution of a few tens of
microarcseconds.

At optical wavelengths, achieving microarcsec resolution requires a
telescope (or an interferometer) larger than 200~m, which is still
unattainable. Recent technological advances at millimeter wavelengths,
however, have allowed combining all available millimeter telescopes in
the world into a single globe-sized instrument called the Event
Horizon Telescope (EHT) that aims to take the first ever images of
black holes in the centers of galaxies with horizon scale resolution.
These images will allow us to observe directly the interaction of
matter and magnetic fields with black hole horizons.  Moreover, it
will enable us to resolve strong-field general relativistic phenomena
that have no weak-field counterparts and perhaps test General
Relativity itself.

\section{When Observations and Theory Converge}

Building an instrument with a very high resolution at a given
wavelength is not sufficient to take a picture of a black hole. A
number of additional conditions need to be satisfied: the Earth's
atmosphere (for ground-based instruments), the Galaxy (for the Milky
Way black hole), and the accretion flow around the black hole need to
be transparent. It turns out that mm-wavelengths satisfy this trifecta
of conditions, for reasons that are purely coincidental.

In the case of the central black hole in the Milky Way, Sgr~A*, the
transparency of the Galaxy was the first to be studied and verified.
Imaging observations in the 1970s at 3.7~cm and 11~cm
\cite{Balick1974} and, since then, at increasingly shorter radio
wavelengths revealed images with sizes that scale with the square of
the wavelength
$\lambda$~\cite{Davies1976,Lo1985,Bower2004,Bower2006}. These
observations were interpreted in terms of detailed models of
interstellar scattering~\cite{Backer1978,Narayan1989,Goodman1989} and
within the context of additional observations of the scattering screen
towards the galactic center~\cite{vanLang1992,Lazio1998a,Lazio1998b} as
being dominated by the blurring from free electrons in the
Galaxy. Extrapolating the expected and observationally verified
$\lambda^2$ dependence of the blurring effect to shorter wavelength
led to the conclusion that they will become negligible at mm
wavelengths. Later observations at even shorter wavelengths confirmed
this by detecting image sizes that deviate from the $\lambda^2$ law, a
result interpreted as caused by resolving the intrinsic size of the
accretion flow image~\cite{Lo1993,Krichbaum1998,Shen2005,Bower2006}.

The transparency of the main bulk of the accretion flow was not
explored until the mid-1990s. Early attempts to simulate the
observational appearance of an accretion flow around a black hole
focused on geometrically thin, optically thick accretion disks (these
disks are often referred to as Shakura-Sunyaev~\cite{Shakura1973} or
Novikov-Thorne disks~\cite{Novikov1973}). Following the work of
Refs.~\cite{Narayan1994,Narayan1995}, however, it became apparent
that, at the low inferred accretion rates of most supermassive black
holes in our vicinity, their accretion flows are mostly geometrically
thick and optically thin, with the dissipated energy getting primarily
advected into the black holes. Soon afterward, the so-called ADAF
(advection dominated accretion flows) or, more generally, RIAF
(radiatively inefficient accretion flows) models were shown to be
consistent with the spectral observations of
Sgr~A*~\cite{Narayan1995a} as well as of other supermassive black
holes, such as the one in center of the M87
galaxy~\cite{DiMatteo2000}. One of the key predictions of these
radiatively inefficient flows (shared also with other models) is the
fact that the radio to IR spectra are dominated by self-absorbed
synchrotron emission, with the synchrotron photosphere shrinking
rapidly with decreasing wavelength.

The parallel efforts that led to the observational evidence for a
shrinking size of the image of Sgr~A* and to the theoretical
predictions of an increasingly more transparent accretion flow with
decreasing wavelength converged at the dawn of the new millennium.
Ref.~\cite{Falcke1998} modeled the observed intrinsic image size of
Sgr~A* as a function of wavelength in terms of resolving an
increasingly more compact self-absorbed emitting region and concluded
that, if this trend continues, the size at millimeter wavelengths will
be comparable to that of the black-hole horizon. In the context of
ADAF models of the accretion flow, Ref.~\cite{Ozel2000} calculated the
predicted size of the synchrotron photosphere as a function of
wavelength and showed that it was consistent with the then-available
observations of Sgr~A* and that the accretion flow would become fully
transparent all the way to the horizon at mm-to-IR wavelengths.

The next issue to address, if we were to observe the transparent
accretion flow around a black hole with horizon-scale resolution, is
what would the signature of the presence of the horizon be. As early
as the 1920's, von Laue calculated the cross section of a
Schwarzschild black hole to a parallel beam of photons arriving from
infinity and found it to have a radius equal to $\sqrt{27} GM/c^2$,
where $G$ is the gravitational constant, $M$ is the black-hole mass,
and $c$ is the speed of light.  In 1973, soon after the discovery of
the Kerr metric, Bardeen~\cite{Bardeen1973} generalized the
calculation of the cross section to a spinning black hole. Motivated
by Bardeen's work, Ref.~\cite{Luminet1979} calculated horizon-scale
images of general geometrically thin disks using very early
computational algorithms and hand-drawn (!)  visualizations of the
results. As computers became more powerful and General Relativistic
radiative transfer algorithms became available, so did simulations of
the observational appearance of Sgr~A* at horizon scales.
Ref.~\cite{Falcke2000} calculated horizon-scale images for Sgr~A*
using simple profiles for the electron emissivity in the vicinity of
the horizon.  In that paper, they also coined the term ``black-hole
shadow'' to refer to the silhouette that a black-hole will cast on the
beam of photons. Later on, Ref.~\cite{Broderick2006} calculated
horizon-scale images specifically of radiatively inefficient accretion
flow models that were shown to agree with all other spectral
observations of this source.

The parallel development of theoretical and observational work
continued~\cite{Krichbaum1993,Backer1993,Rogers1994,Krichbaum1998}
and, in 2008, Doeleman and collaborators made a successful
interferometric observation of Sgr~A* at 1.3 mm, using an array
comprising only three sites, in Arizona, in California, and in
Hawaii~\cite{Doeleman2008}. Even though these observations had a very
limited baseline coverage to allow imaging of the source, they
nevertheless demonstrated that the scale of the image was smaller than
$\simeq 5$ Schwarzschild radii, consistent with the expected size of
the black-hole shadow. A similar experiment in 2012 with the black
hole in the center of M87 gave comparable
results~\cite{Doeleman2012}. Both of these observations provided the
proof of principle for interferometric imaging of horizon-scale
structures in Sgr~A* and M87 at millimeter wavelengths and the impetus
for the construction of the Event Horizon Telescope.

\section{The Event Horizon Telescope}

The Event Horizon Telescope (EHT) is a globe-sized array of radio
telescopes operating currently at 1.3~mm that aims to capture some of
the highest-resolution astronomical images ever made, including the
first images of astrophysical black holes with horizon-scale
resolution~\cite{Doeleman2009a}. In its 2018 array configuration, the
EHT involved nine stations around the globe: from Hawaii to France and
from Greenland to the South Pole.

\begin{figure*}
\centerline{  \includegraphics[width=0.99\textwidth]{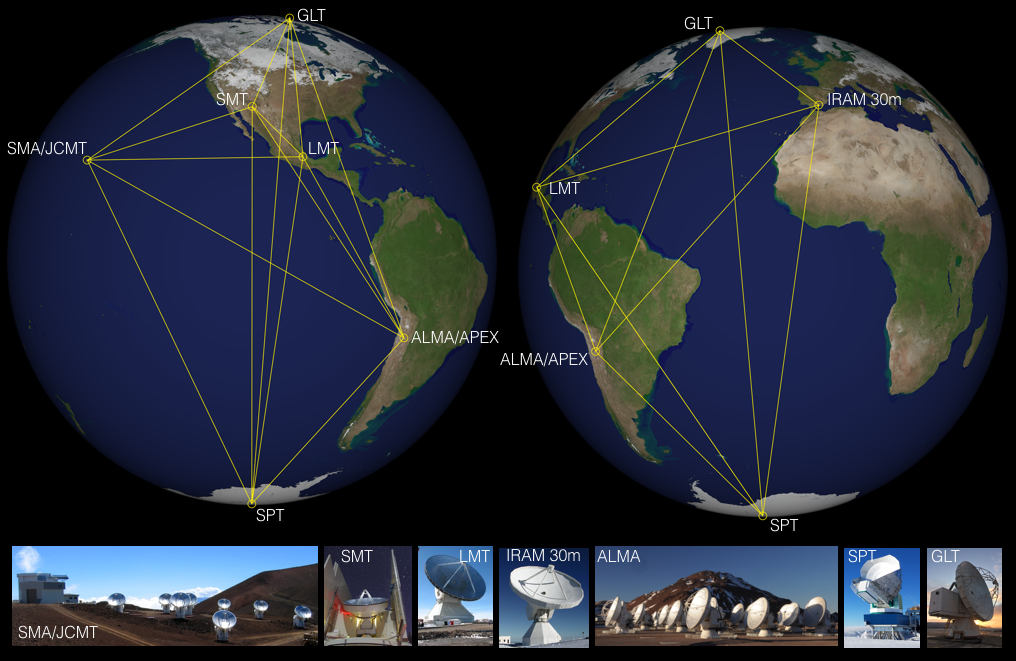}}
\caption{The mm telescopes around the world that comprise
the Event Horizon Telescope. (Credit: D.\ Marrone/UofA)}
\label{fig:eht}       
\end{figure*}

Early observations with a subset of the array have demonstrated the
feasibility of the EHT project. In particular, observations with small
subsets of the full array have shown: {\em (i)} horizon-scale
structures at 1.3~mm for Sgr~A*~\cite{Doeleman2008}; {\em (ii)} source
substructure and variability~\cite{Fish2016}; {\em (iii)\/} highly
polarized emission at horizon scales, indicating the presence of large
scale magnetic fields in the vicinity of the black
hole~\cite{Johnson15}; and {\em (iv)} horizon-scale structure at 1.3~mm
for the black hole in the center of M87~\cite{Doeleman2012}.

\begin{figure}
\includegraphics[width=0.7\textwidth]{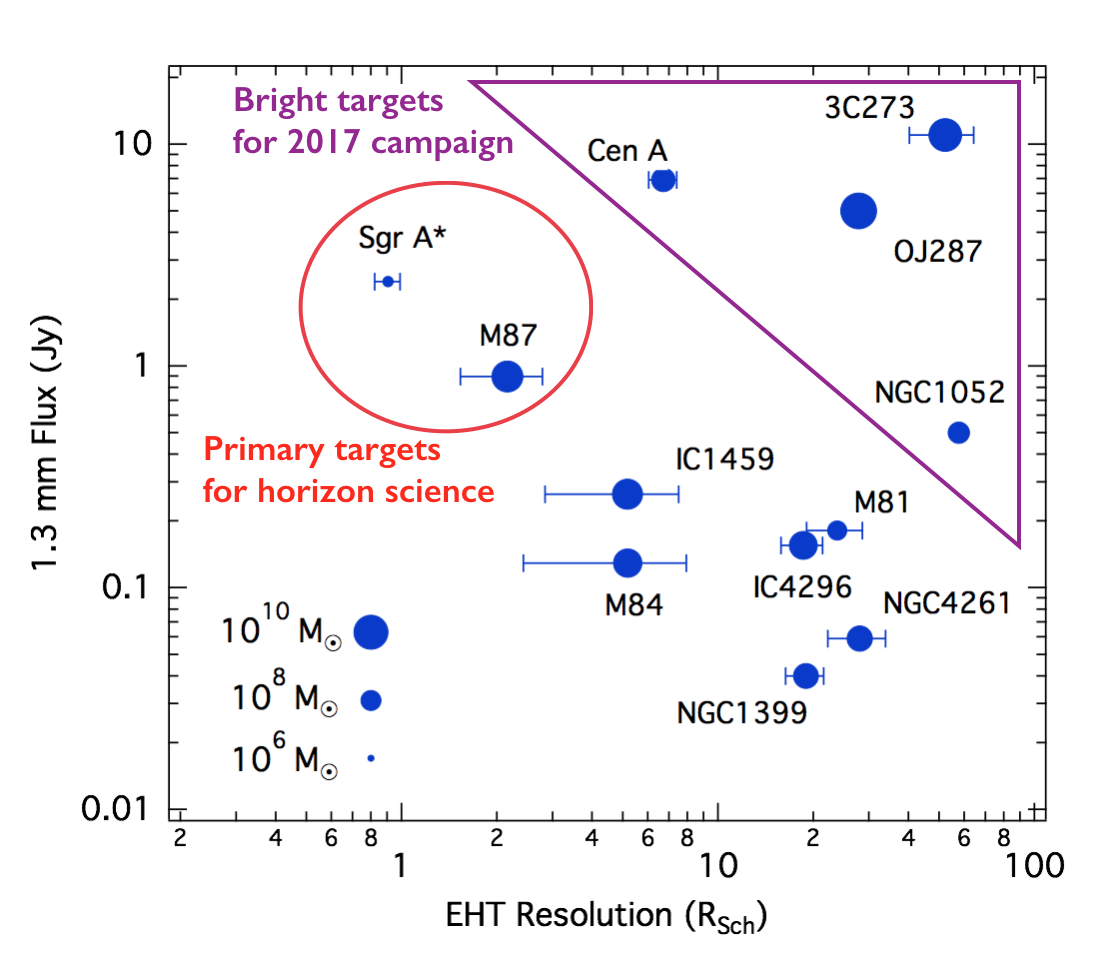}
\caption{Prime targets for observations with the Event Horizon
  Telescope. The 1.3mm flux and nominal EHT resolution in units of the
  Schwarzschild radius is shown for a number of known black-hole
  targets. The two primary targets for horizon-scale science,
  Sgr~A*\ and M87, as well as four additional bright targets that were
  observed during the April 2017 campaign are indicated. The source
  3C279, which is also a science target for the array, lies outside
  the boundaries of this plot.}
\label{fig:sources}       
\end{figure}

The EHT has since undergone a phase of rapid development and
expansion. The first full-array observations were performed in
April~2017, with six targets~(see Fig.~\ref{fig:sources}). In the case
of its two primary targets, Sgr~A*\ and M87, the EHT is designed to
achieve horizon-scale resolution. 

In order to exploit the new opportunities that the EHT offers, novel
algorithms have been developed to reconstruct images from the
interferometric
data~\cite{Honma2014,Bouman2016,Chael2016,Akiyama2017,Bouman2017,Johnson2017,Chael2018},
correct for the blurring effects of interstellar
scattering~\cite{Fish2014,Johnson2015a,Johnson2016a,Johnson2016b},
quantify the coherence lengths of magnetic fields near the
horizons~\cite{Gold2017}, search for signatures of the black-hole
shadows, and perform tests of general
relativity~\cite{Johannsen2010a,Broderick2014,Psaltis2015b,Psaltis2016}. Furthermore,
new tools have been implemented to perform Bayesian comparisons of the
imaging and timing data to the predictions of semi-analytic models of
accretion flows~\cite{Broderick2009}, of geometric models of
structures~\cite{Kamruddin2013,Benkevitch2016}, and of time-dependent
General Relativistic MagnetoHydroDynamic (GRMHD)
simulations~\cite{Kim2016}. Finally, these analysis methods are being
validated using large suites of state-of-the-art GRMHD simulations
with parameters tuned to match existing observations
~\cite{Moscibrodzka2009,Moscibrodzka2014,Dexter2009,Dexter2010,Chan2015,Ressler2015,Gold2017,Porth2016,Ressler2017,Medeiros2016a,Medeiros2016b,Moscibrodzka2016,Moscibrodzka2017}.
  
The expected images of the shadows cast by the black holes on their
surrounding emission, polarimetric maps, and time-variability studies
are poised to open novel ways of observing and understanding
astrophysical black holes. Even though a lot will be learned about the
interaction of the back holes with the plasma in the accretion flows
and in the jets that surrounds their horizons, this review will focus
entirely on the prospect of probing and testing General Relativity
with EHT observations.


\section{Why Test General Relativity with the EHT?}

During the last century, many General Relativistic (GR) predictions
have been tested in various astrophysical settings. A large number of
these tests have been performed within the Solar System, primarily
because of the opportunity such tests offer to control systematic
uncertainties~\cite{Will2017}. More recently, quantitative tests have placed
significant constraints on potential modifications of GR at vastly
different scales, from the strongest gravitational fields of neutron
stars and stellar-mass black holes~\cite{Psaltis2008,Berti2015} to the weakest
gravitational fields probed by cosmological phenomena~\cite{Koyama2016}.

It is often said that GR has passed all these tests with flying
colors. This begs the question, then, of why we care to continue
testing this theory with black holes, especially in the era of
gravitational wave detection from coalescing compact objects with
LIGO/VIRGO~\cite{Abbott2016}. There are several motivations for testing GR with
the EHT, which I discuss in some detail below; some are primarily
empirical and others are mostly theoretical.

First, black-hole spacetimes are qualitatively different than those of
other astrophysical objects in terms of testing the underlying theory
of gravity. These are the spacetimes in which GR predicts its own
demise by forming singularities at their very centers. Moreover, these
are the spacetimes that lead to unexpected paradoxes, such as the
information paradox~\cite{Hooft2016} and
firewalls~\cite{Almheiri2012}, when one tries to perform calculations
with quantum fields in the vicinities of their horizons. Perhaps it is
true that all singularities are clothed behind horizons, as the cosmic
censorship conjecture postulates. Perhaps the resolution to the
information paradox occurs at scales that are too small to be detected
observationally. It is quite plausible, however, that the quantum
structure of black holes will leave classical, horizon-scale
signatures that will be discernible in gravitational tests with black
holes~\cite{Giddings2017a}. Observations of black holes that resolve
horizon scales (either in the electromagnetic spectrum or with
gravitational waves) will allow us to test the cosmic censorship
hypothesis, measure the properties of their spacetimes, and look for
signatures of quantum structures.

Second, the horizon-scale images of the supermassive black holes that
will be observed with the EHT probe gravitational fields that are
vastly different than those probed in all other GR tests, with or
without black holes. Ref.~\cite{Baker2015} quantified the strength of
the gravitational fields probed in different astrophysical systems and
by different experiments in terms of the typical magnitude of the
gravitational potential and the gravitational curvature in each
setting. For a distance $r$ away from a point mass $M$, the
gravitational potential and curvature are of order
\begin{equation}
  \epsilon\equiv \frac{GM}{rc^2}
\end{equation}
and
\begin{equation}
  \xi\equiv \frac{GM}{r^3 c^2}\;,
\end{equation}
respectively. (For more precise definitions that are based on invariant
quantities and are more broadly applicable, see the discussion in
Ref.~\cite{Baker2015}.)

\begin{figure*}
  \centerline{  \includegraphics[width=0.5\textwidth]{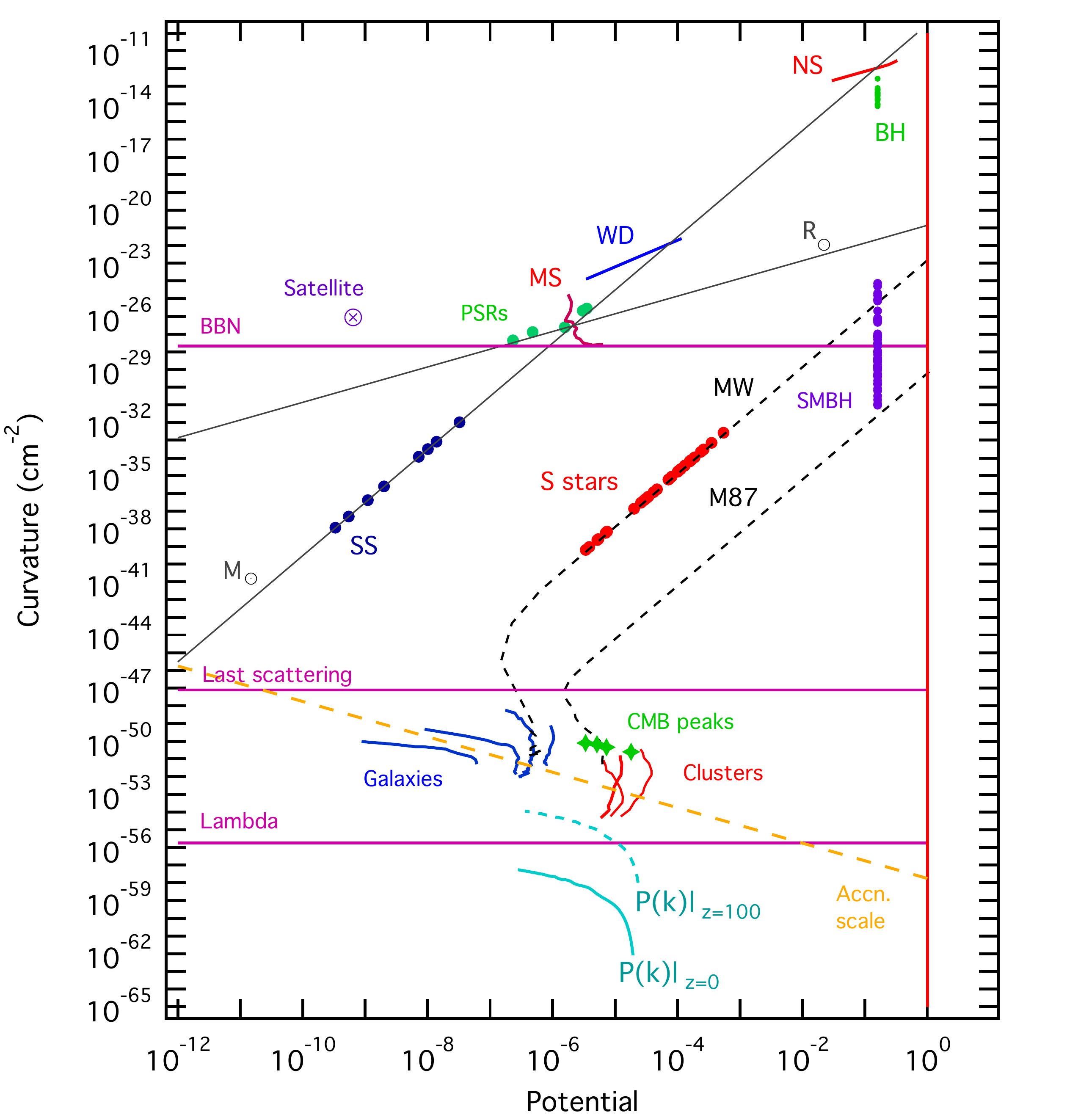}
  \includegraphics[width=0.5\textwidth]{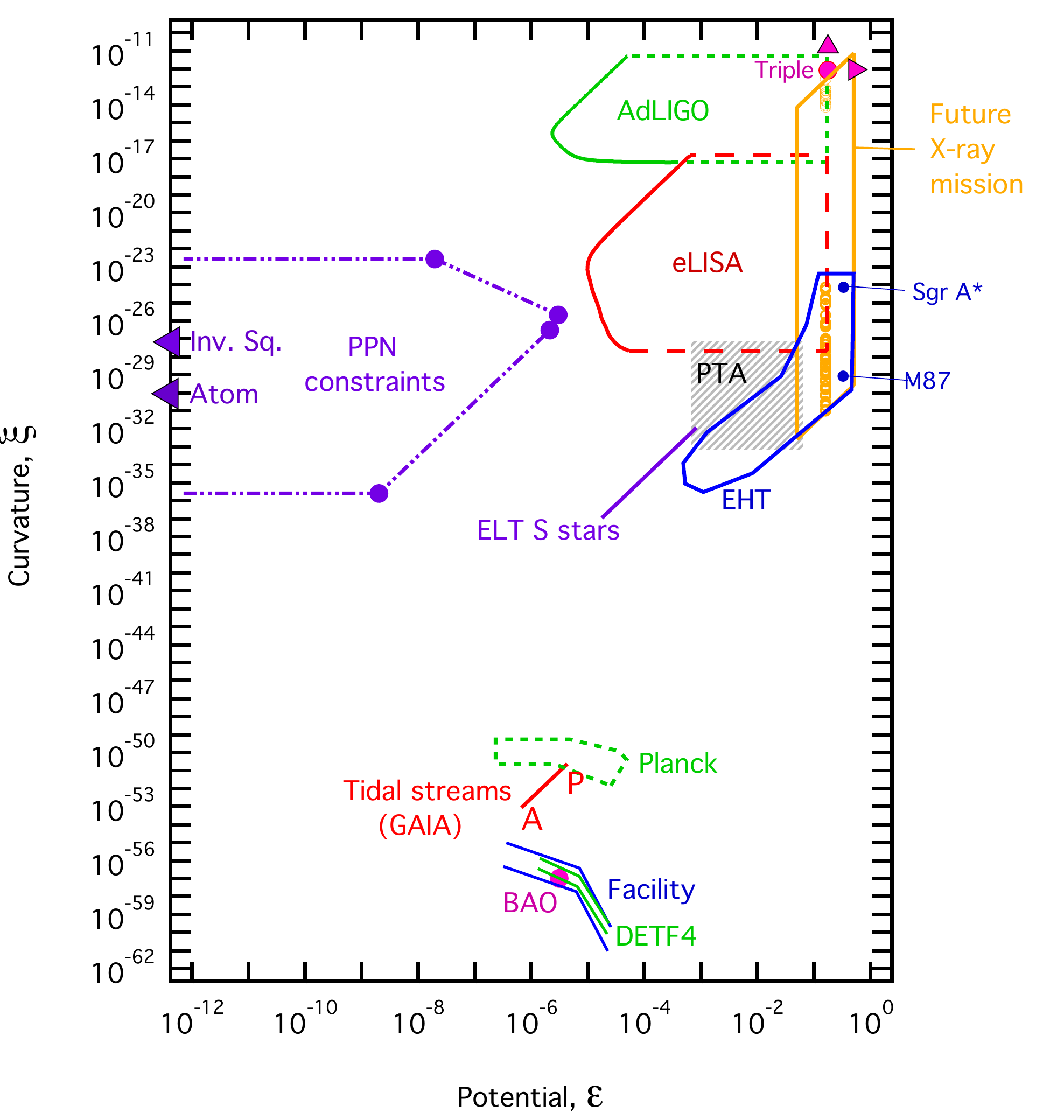}  }
  \caption{The gravitational potential and curvature probed {\em
      (left)\/} in different astrophysical settings and {\em
      (right)\/} with different types of observations. The Event
    Horizon Telescope (EHT) offers the possibility of gravitational
    tests that are complementary to those of other current and planned
    experiments, such as those with LIGO/VIRGO, LISA, Pulsar Timing
    Arrays (PTA), and with optical observations of S-Stars around
    \sgra. (After Ref.~\cite{Baker2015}.)}
\label{fig:fields}       
\end{figure*}

When seen in the parameter space defined by these two quantities
(Fig.~\ref{fig:fields}), it becomes clear that the EHT will probe
gravitational potentials that are $\sim 5$ orders of magnitude higher
than those of solar system tests and gravitational curvatures that are
$\sim 15-20$ orders of magnitude different than those probed by
LIGO/VIRGO and by cosmological tests. Even though, in GR, there is no
characteristic scale in curvature and, therefore, all astrophysical
settings with comparable potentials provide similar constraints, this
is not necessarily the case for theories that deviate from GR. As an
example, settings with the same potential but different curvatures
lead to different gravitational constraints in a large number of
theories with screening mechanisms~\cite{Joyce2015}. In the context of
tests with compact objects, Ref.~\cite{Yagi2016} showed how black-hole
and neutron-star binaries, both of which harbor compact objects of
comparable potentials and curvatures, give rise to vastly different
constraints on quadratic gravity. Another example that demonstrates
the utility of complementary tests at different scales is the recent
fifth-force tests with S-stars reported in Ref.~\cite{Hees2017}.
Together with terrestrial (e.g., LIGO/VIRGO, S~Star monitoring) and
space-based (e.g., LISA) observations, the EHT will provide a
complementary and comprehensive survey of near-horizon gravitational
effects with black holes at curvatures that span over 25 orders of
magnitude.

Finally, EHT and LIGO/VIRGO tests of GR probe different (albeit highly
connected) aspects of the gravitational theory.  In GR tests with the
EHT, one uses effectively test particles (photons and plasma) to
probe the properties of the stationary spacetimes of black holes at
very long times after their formation. Because of this, tests with the
EHT are actually metric tests and are agnostic with respect to the
underlying theory of gravity (as long as it obeys the equivalence
principle). In contrast, gravitational waves test the dynamics of the
theory during violent merger events and can be used to infer the
properties of the stationary metrics only via their dynamics (see the
discussion in Ref.~\cite{Hughes2006}). This complementarity is
important in tests that involve black holes because a large number of
gravitational theories share the exact same black-hole solutions with
GR but they differ in the predicted dynamics and gravitational wave
signatures~\cite{Psaltis2008a,Barausse2008,Sotiriou2012,Berti2015}.
As an example, the shapes and sizes of the shadows of black holes,
which the EHT aims to observe, are very sensitive to the quadrupole
and higher moments of the stationary black hole spacetimes (see
discussion below). For this reason, EHT observations have the
potential of testing the no-hair theorem with astrophysical black
holes. On the other hand, LIGO/VIRGO observations of gravitational
waves measure the time evolution of the quadrupole moments during
coalescence but are not currently sensitive enough to measure the
moments of the stationary spacetimes during the ring-down
phases and test the no-hair theorem~\cite{Berti2016}.

With this motivation in mind, I will now turn into discussing the
various gravity tests with EHT observations that have been proposed.

\section{Tests with Black-Hole Shadows}

\begin{figure*}
  \centerline{ \includegraphics[width=0.33\textwidth,height=0.33\textwidth]{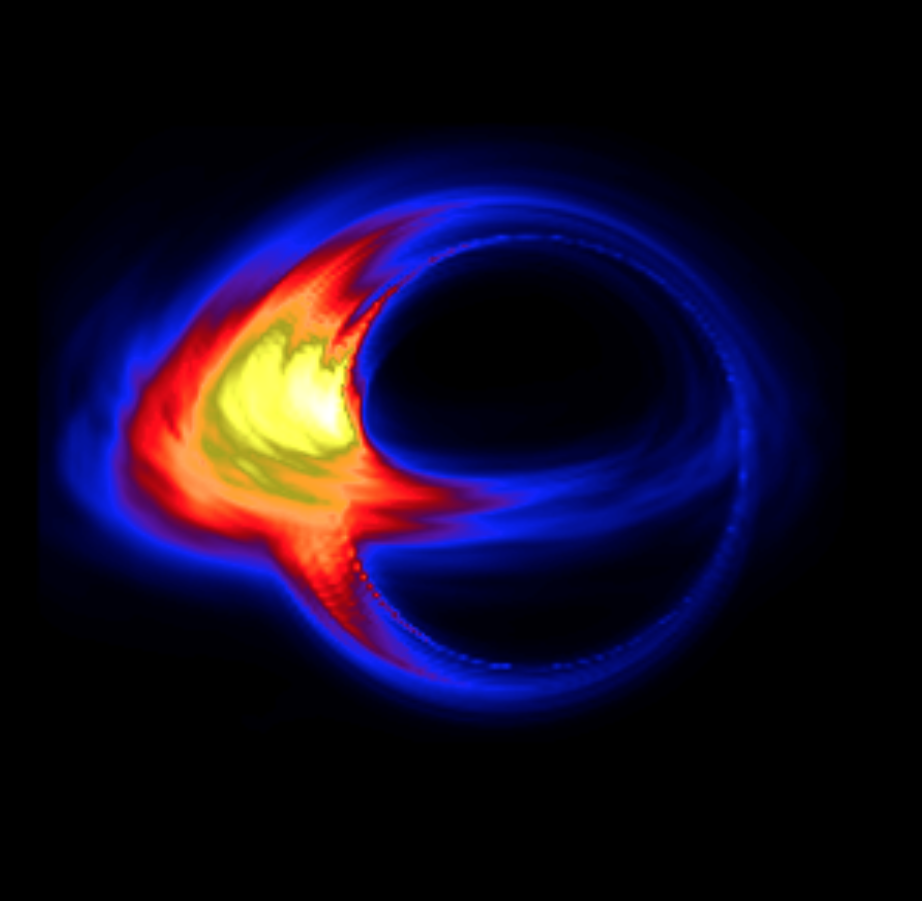}
    \includegraphics[width=0.33\textwidth,height=0.33\textwidth]{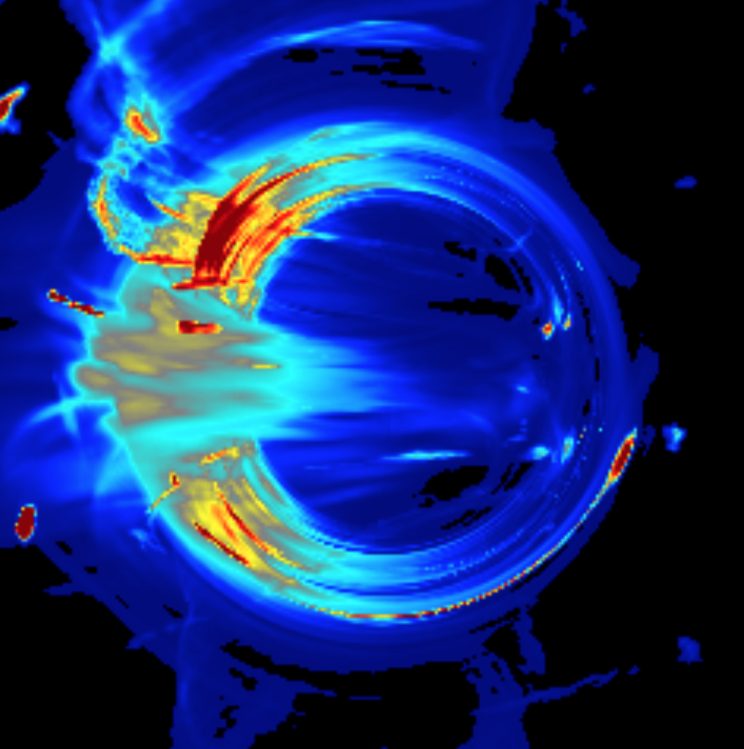}
    \includegraphics[width=0.33\textwidth,height=0.33\textwidth]{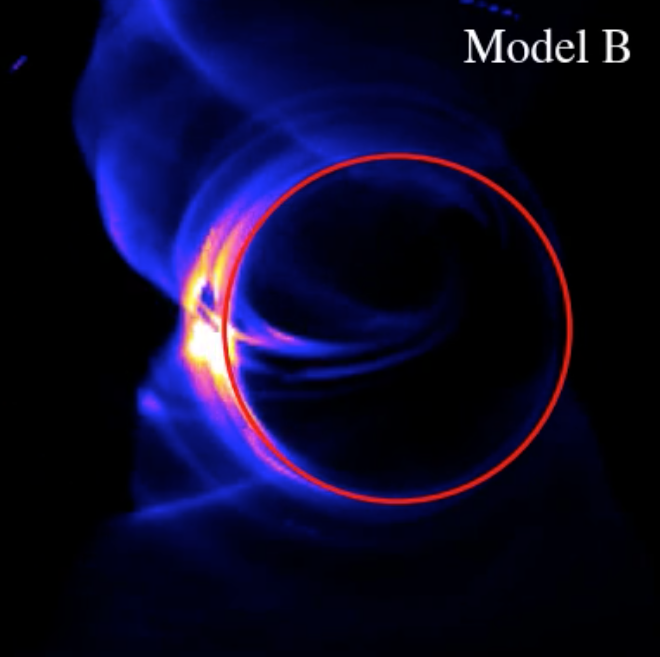} }
  \caption{Predicted 1.3mm images for \sgra\ from three different
    GRMHD
    simulations~\cite{Dexter2009,Moscibrodzka2014,Medeiros2016a}.
    Even though the simulations employ different algorithms and
    different prescriptions for the sub-grid plasma physics, they all
    show prominent features at the outline of the black-hole shadow
    (marked by a red circle in the rightmost panel). The size of the
    black-hole shadow is $\sim 10GM/c^2$.}
\label{fig:grmhd}       
\end{figure*}

EHT observations of the two primary targets, \sgra\ and M87, aim to
generate the first images of the shadows the black holes cast on the
surrounding emission. Figure~\ref{fig:grmhd} shows a compilation of
different images at 1.3~mm that are representative of large suites of
GRMHD simulations using different algorithms and performed by
different researchers. The detailed emission structures are different
between images, depending on the way that various simulations treat
the initial conditions, the heating of the electrons in the plasma,
and the spin and orientation of the black hole with respect to the
observer. All images, however, are characterized by a prominent
black-hole shadow, which is only partially obscured by intervening
plasma orbiting on the equatorial plane.

\subsection{Properties of Black-Hole Shadows}

The outline of a black-hole shadow is determined entirely by the
location of the photon orbits and by gravitational lensing. All
photons that cross a photon orbit with a momentum vector that point
inwards eventually also cross the horizon (in the absence of any
additional interaction with plasma) and do not reach distant
observers. Because, in an accretion flow, most emission takes place
outside the radius of the innermost stable circular orbit (ISCO),
which itself is outside the radius of the photon orbit, it follows
that most photons that cross the radius of the photon orbit will have
inward momenta and will eventually disappear behind the
horizon. Therefore, the black hole casts a shadow on the surrounding
emission with a size and shape determined by the location of the
various photon orbits at different orientations with respect to the
black-hole spin axis. The outline of the shadow corresponds to the
impact parameters for the trajectories of photons that have barely
grazed the corresponding photon orbit. However, its size, as measured
by a distant observer, will be larger than the projected radius of the
photon orbit, because of gravitational lensing.

\begin{figure*}
  \centerline{ \includegraphics[width=0.5\textwidth]{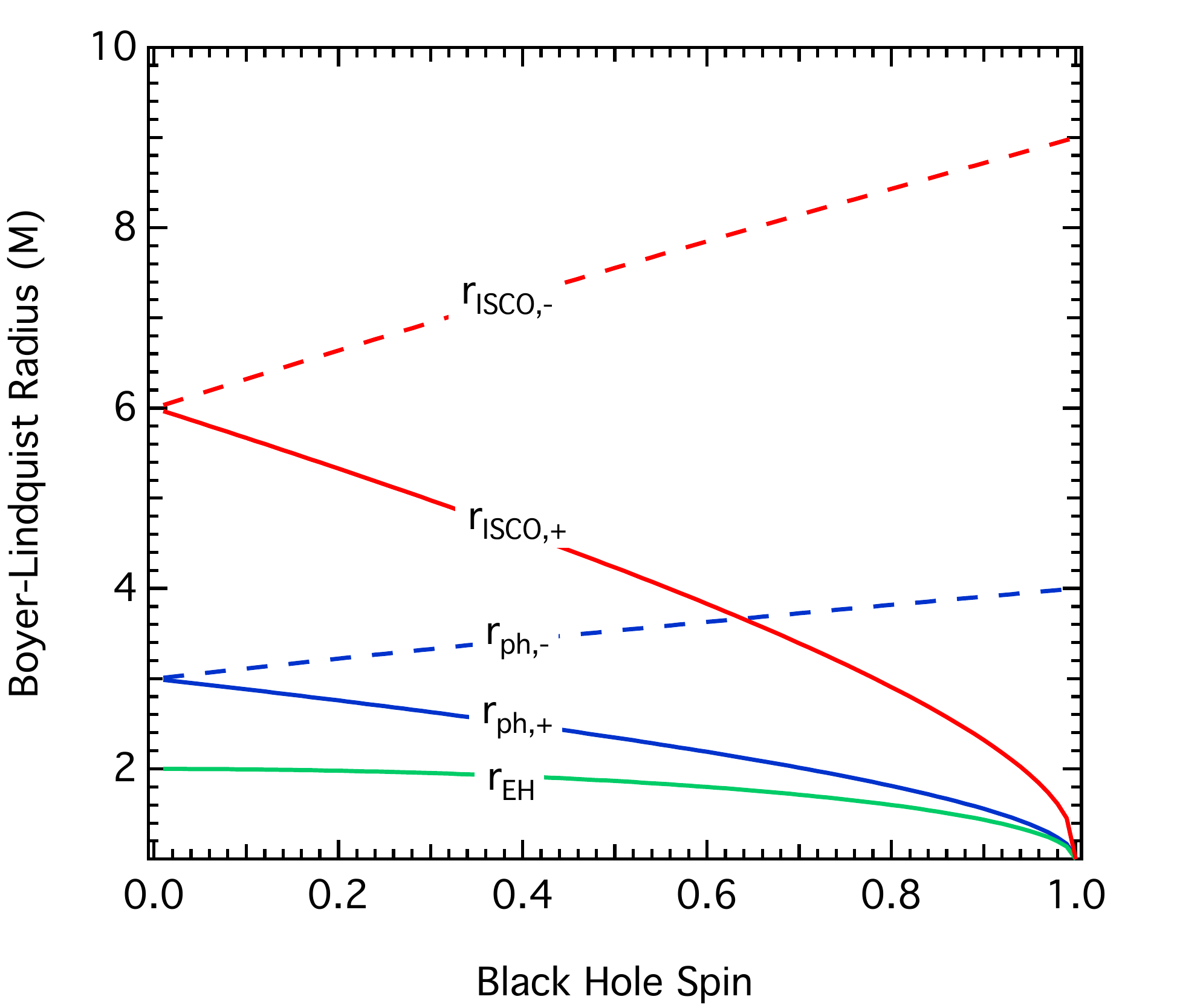}
    \includegraphics[width=0.5\textwidth,height=0.46\textwidth]{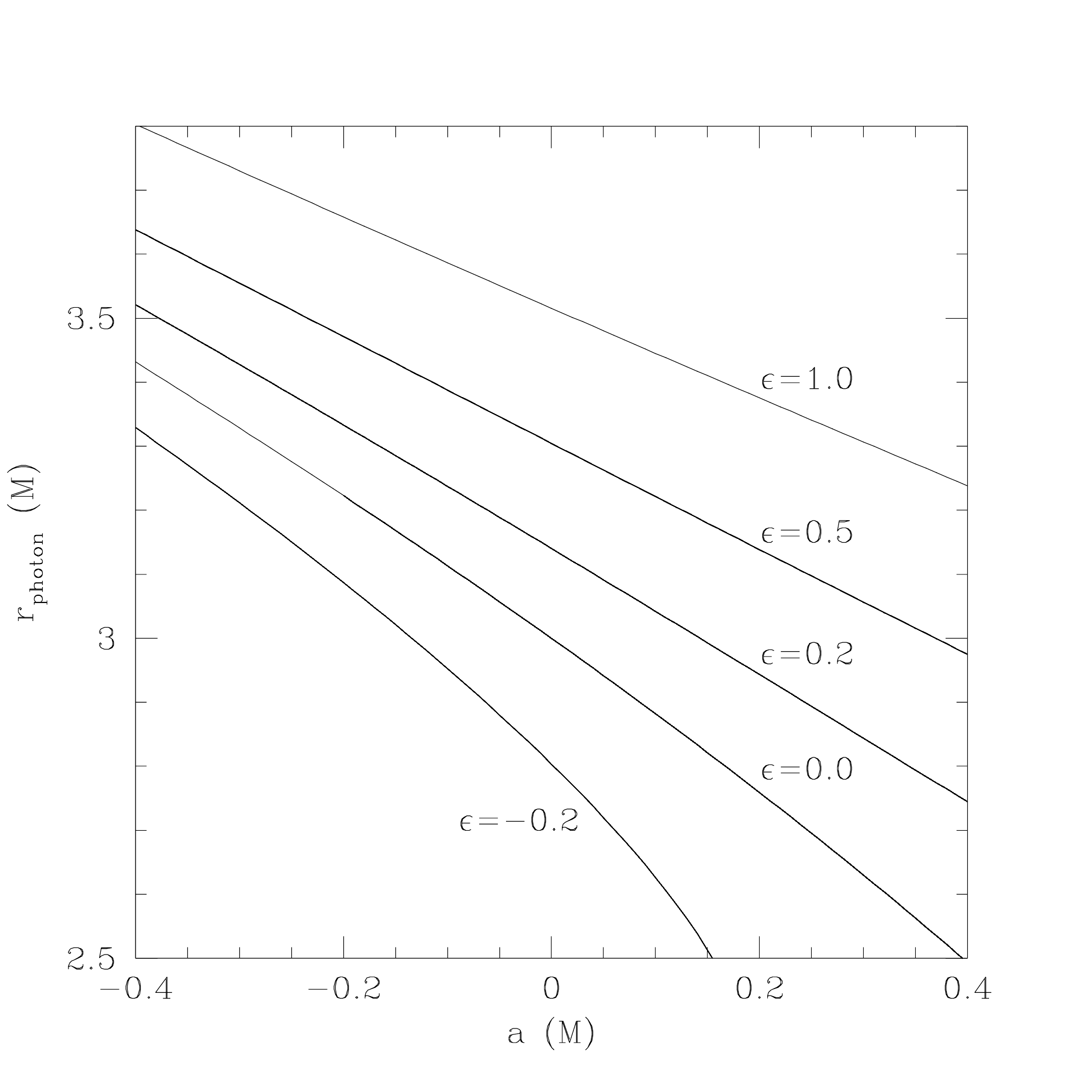}
  }
  \caption{{\em (Left)\/} Characteristic radii in the spacetimes of
    Kerr black holes, as a function of the black-hole spin. Shown are
    the Boyer-Lindquist radii of the event horizon ($r_{\rm EH}$), of
    the prograde and retrograde photon orbits ($r_{\rm ph,\pm}$), and
    of the prograde and retrograde equatorial ISCO ($r_{\rm
      ISCO,\pm}$). (After Ref.~\cite{Bardeen1973}.) {\em (Right)\/}
    The dependence of the radius of the prograde photon orbit on
    black-hole spin for spacetimes that violate the no-hair
    theorem~\cite{Johannsen2010}. The parameter $\epsilon$ measures
    the deviation of the quadropole moment of the spacetime from the
    Kerr value.}
\label{fig:radii}       
\end{figure*}

For a non-spinning black hole, the Schwarzschild radius of the photon
orbit is independent of orientation and equal to
\begin{equation}
r_{\rm ph}=\frac{3GM}{c^2}\;.
\end{equation}
The effect of gravitational lensing is to magnify this to a shadow size of
\begin{equation}
  R_{\rm shadow}=\frac{\sqrt{27}GM}{c^2}\;.
\end{equation}
For a spinning black hole, the presence of non-zero multipole
mass moments breaks the spherical symmetry of the
spacetime. Nevertheless, photon orbits with constant Boyer-Lindquist
radii still exist but with distances from the black hole that depend
on the orientation of the orbital angular momentum~\cite{Teo2003}. The
existence of closed, spherical photon orbits in a Kerr spacetime is
intimately related to the existence of the Carter
constant~\cite{Cunha2017}. Because of frame dragging, the radius of a
photon orbit depends on the relative orientation of the orbital
angular momentum with respect to the black-hole spin. For equatorial
orbits, the Boyer-Lindquist radii of the prograde ($+$) and of the
retrograde photon orbits ($-$) are given by~\cite{Bardeen1972}
\begin{equation}
  r_{\rm ph,\pm}=\frac{2GM}{c^2}\left\{1+\cos\left[\frac{2}{3}\arccos\left(
    \mp\vert a\vert\right)\right]\right\}\;,
\end{equation}
where $0<\vert a\vert\le 1$ is the specific angular momentum of the
black hole per unit mass. Figure~\ref{fig:radii} shows the
Boyer-Lindquist radius of prograde and retrograde photons orbits at
different black-hole spins and compares them to the radii of the ISCO
and of the event horizon. The multipole mass moments of the spacetime
and frame dragging affect also the degree of gravitational lensing
that photons experience on their way to a distant observer. The net
effect on the outline of the black-hole shadow can still be calculated
analytically in the parametric
form~\cite{Bardeen1973,Chandrasekhar1983,Takahashi2004,Bozza2006,Psaltis2015b}
\begin{eqnarray}
\alpha(r)&=&-
\frac{\left[a^2 (r+1)+(r-3) r^2\right] \csc i}{a (r-1)}
\left(\frac{GM}{c^2}\right)
\nonumber\\
\beta_\pm(r)&=&\pm\frac{1}{a(r-1)}
\left\{a^4(r-1)^2 \cos^2 i-\left[a^2 (r+1)+(r-3) r^2\right]^2 \cot^2 i
\right.\nonumber\\
&&\qquad\qquad\left.
-r^3 \left[(r-3)^2 r-4 a^2\right]\right\}^{1/2} \left(\frac{GM}{c^2}\right)
  \label{eq:shadow}
 \end{eqnarray}
Here, $i$ is the inclination of the observer measured
from the spin axis of the black hole, the parameter $r$ takes values
in an interval bounded by $[r_{\rm ph,-},r_{\rm ph,+}]$ such that
$\alpha$ and $\beta$ are real numbers, and $\alpha$ and $\beta$ are
two orthogonal angular coordinates on the image plane of a distant
observer with $\alpha$ perpendicular to the spin axis of the black
hole.

\begin{figure*}
  \centerline{ \includegraphics[width=0.5\textwidth]{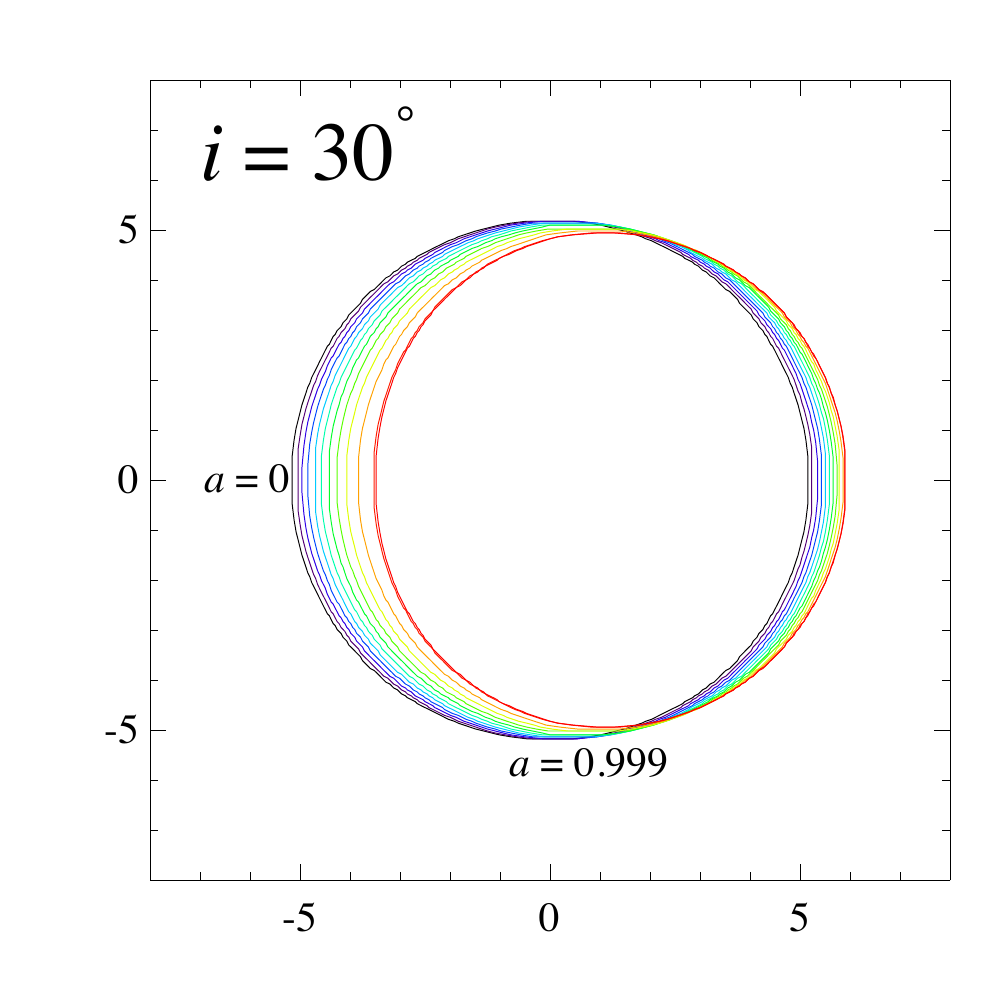}
    \includegraphics[width=0.5\textwidth]{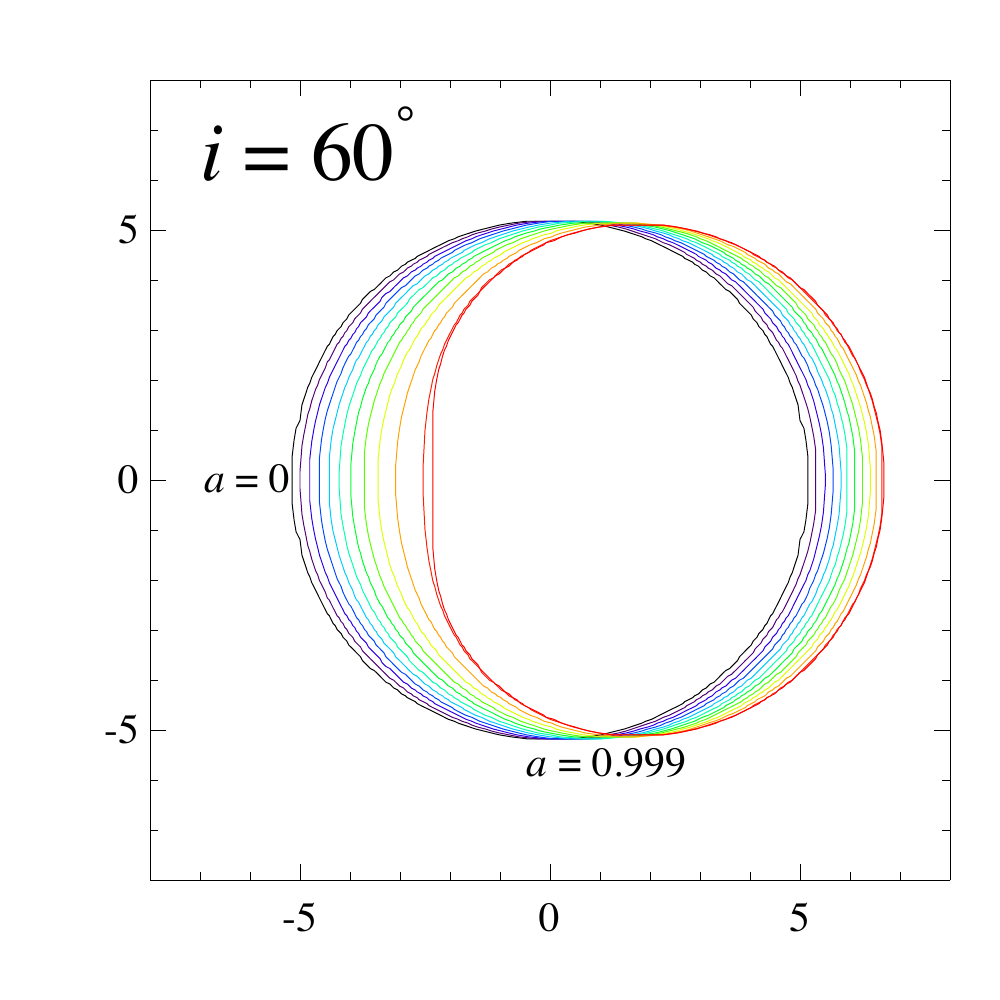} }
  \caption{The outline of the shadow cast by a Kerr black hole, for
    different values of the black-hole spin and the inclination of the
    observer. For all but the highest spins and for all but the
    highest inclinations, the shadow remains nearly circular, with a
    radius that depends very weakly on spin or inclination. }
\label{fig:shadow}       
\end{figure*}

\begin{figure*}
  \centerline{ \includegraphics[width=0.99\textwidth]{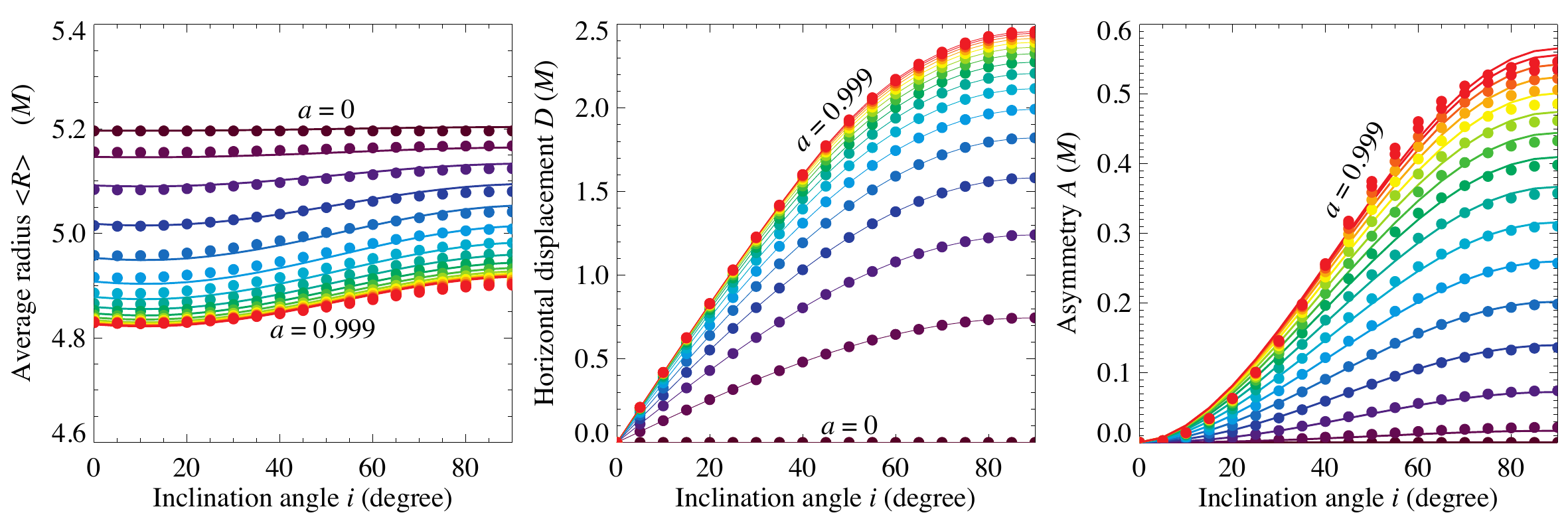}}
  \caption{The average radius, horizontal displacement, and asymmetry
    of the shadow of a Kerr black hole, for different black-hole spins and
  observer inclinations~\cite{Chan2013}. }
\label{fig:properties}       
\end{figure*}

Figure~\ref{fig:shadow} shows the outline of a black-hole shadow for
different spins and different inclinations of the observer. There are
a number of immediate results that can been seen from this
picture:

\noindent {\em (i) The size of the shadow of a Kerr black
  hole depends extremely weakly on spin and inclination.\/} We can
  calculate the half opening angle (``radius'') of the shadow on the
  spin equator for $i=\pi/2$ as
\begin{eqnarray}
  R_{\rm shadow, eq}&=&\frac{1}{2}\left[\alpha(r_{\rm ph, +})
  -\alpha(r_{\rm ph,-})\right]\nonumber\\
    &\simeq& \frac{\sqrt{27}GM}{c^2} \left(1-\frac{a^2}{18}\right)
    +{\cal O} (a^4)\;.\label{eq:rsha}
\end{eqnarray}
Equation~(\ref{eq:rsha}) as well as a more systematic study of the
size of the shadow averaged over different directions with respect to
its center~\cite{Johannsen2010a} show that, for all black-hole spins
and observer inclinations (see Fig.\ref{fig:properties}),
\begin{equation}
  \label{eq:shadow_size}
  \langle R_{\rm shadow}\rangle\simeq \frac{5GM}{c^2}\pm 4\%\;.
\end{equation}

\noindent {\em (ii) The center of the black-hole shadow does not
  coincide with the projected position of the origin of the
  spacetime.\/} We can calculate the position of the center of the
black-hole shadow for $i=\pi/2$ as
\begin{eqnarray}
  D=\frac{1}{2}\left[\alpha(r_{\rm ph, +})
  +\alpha(r_{\rm ph,-})\right]\simeq -2a\;.\label{eq:dsha}
\end{eqnarray}
A more systematic study of the size of the shadow averaged over
different directions with respect to its
center~\cite{Bozza2006,Johannsen2010a} show that, for all black-hole
spins and observer inclinations (see Fig.\ref{fig:properties}),
\begin{equation}
  D\simeq -2 a \sin i\;.
\end{equation}

\noindent {\em (iii) The black-hole shadow is nearly circular for all
  but the highest values of its spin.\/} Different definitions and
approximate expressions exist in the literature for the degree of
asymmetry of the
shadow~\cite{Bozza2006,Johannsen2010a,Chan2013,Johannsen2013b}. They
all show a very small degree of asymmetry that, even for the most
rapidly spinning black holes, is $\le$5\% (see
Fig.~\ref{fig:properties}).

It is surprising that, even though all properties of the Kerr
spacetime (e.g., the horizon, the location of the ISCO, the location
of the photon orbit) depend very strongly on the spin of the black
hole (see Fig.~\ref{fig:radii}), the shape and size of its shadow
remains remarkably constant. This has been understood as the result of
the near cancellation of the effects of the spacetime quadrupole and
of frame dragging~\cite{Johannsen2010a}. The quadrupole moment of the
spacetime breaks the spherical symmetry of the problem and would have
caused the black-hole shadow to appear highly elliptical. However,
photons that propagate in the same direction as the black-hole spin
experience the opposite effect of frame dragging along their
trajectories compared to photons that propagate in the opposite
direction. When the spacetime spin $a$ and quadrupole $q$ obey the
Kerr relation, $q=-a^2$, these two effects nearly cancel each other,
causing the black-hole shadow do be nearly circular and displaced.

\subsection{Proposed Tests}

The properties of black-hole shadows that are summarized in
Fig.~\ref{fig:properties} have led to a number of proposed tests of
gravity:

\bigskip

\noindent {\em Cosmic Censorship Tests.---\/} Ref.~\cite{Bambi2009}
proposed a test of the cosmic censorship hypothesis based on using
observations of black-hole shadows to differentiate between Kerr
metrics that are surrounded by horizons (i.e., for $a\le 1$) and those
that are not ($a>1$). This test can be extended to spacetimes of naked
singularities that are not described by the Kerr solution (see, e.g.,
Ref.~\cite{Shaikh2018}). The conclusion of these studies is that the
detection of a black-hole shadow is {\em not\/} a proof of the
presence of an event horizon; naked singularities may also show
shadows, depending on their parameters. When they do not show shadows,
the resulting images are characterized by brightness profiles that are
very centrally peaked. When they do show shadows, their shapes are
often very unusual and their sizes can be significantly different from
those of Kerr black holes. If EHT observations provide conclusive
evidence for the presence of shadows in the images of the primary
targets, then cosmic censorship tests can be performed in parallel
with the null-hypothesis, no-hair theorem, and metric tests described
below, all of which also rely on measuring the shapes and sizes of the
shadows.

\begin{figure*}
  \centerline{ \includegraphics[width=0.5\textwidth]{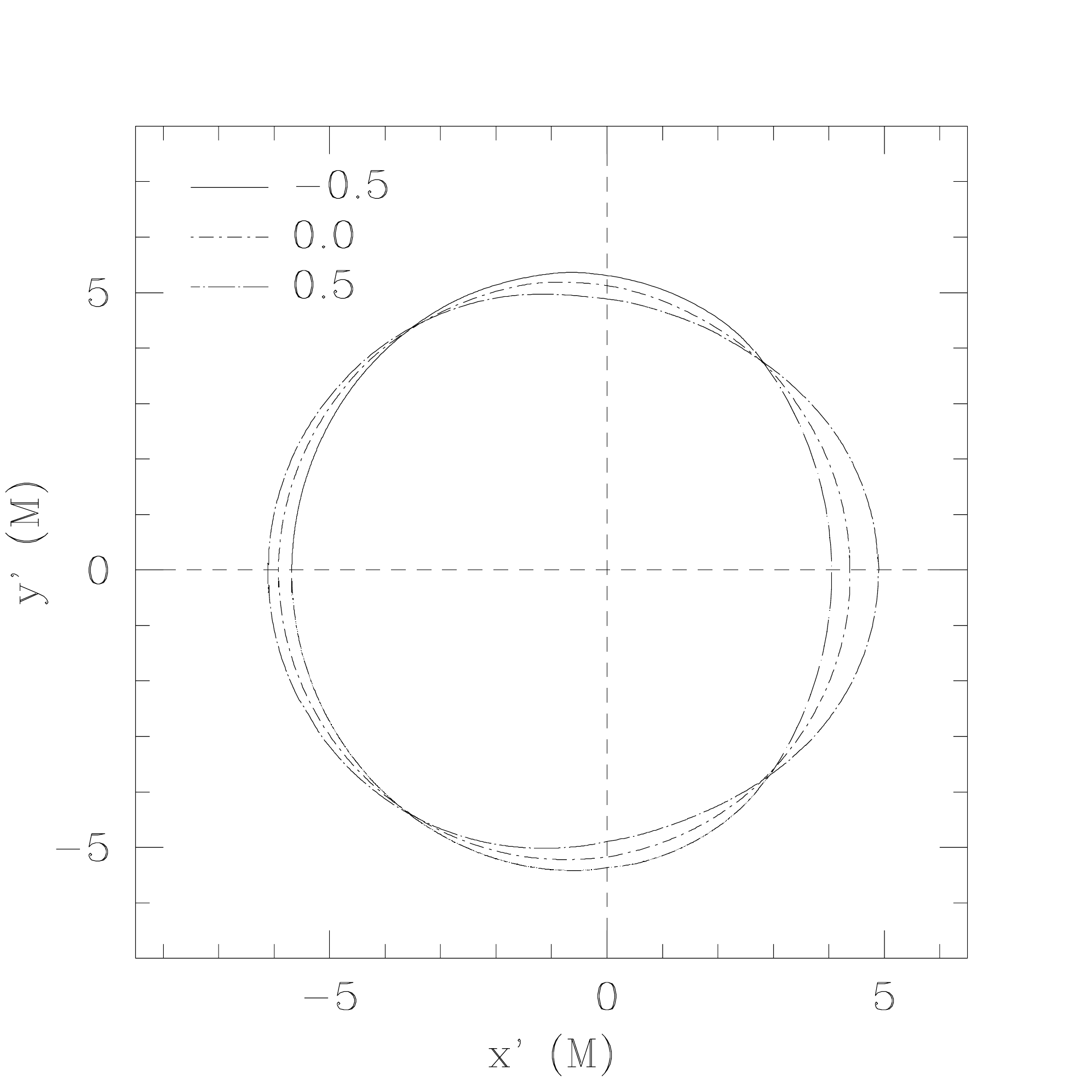}
  }
  \caption{Asymmetric black-hole shadows from quasi-Kerr spacetimes
    that violate the no-hair theorem~\cite{Johannsen2010a}.}
  \label{fig:nonKerrshadows}       
\end{figure*}

\bigskip

\begin{figure*}
  \centerline{ \includegraphics[width=0.5\textwidth]{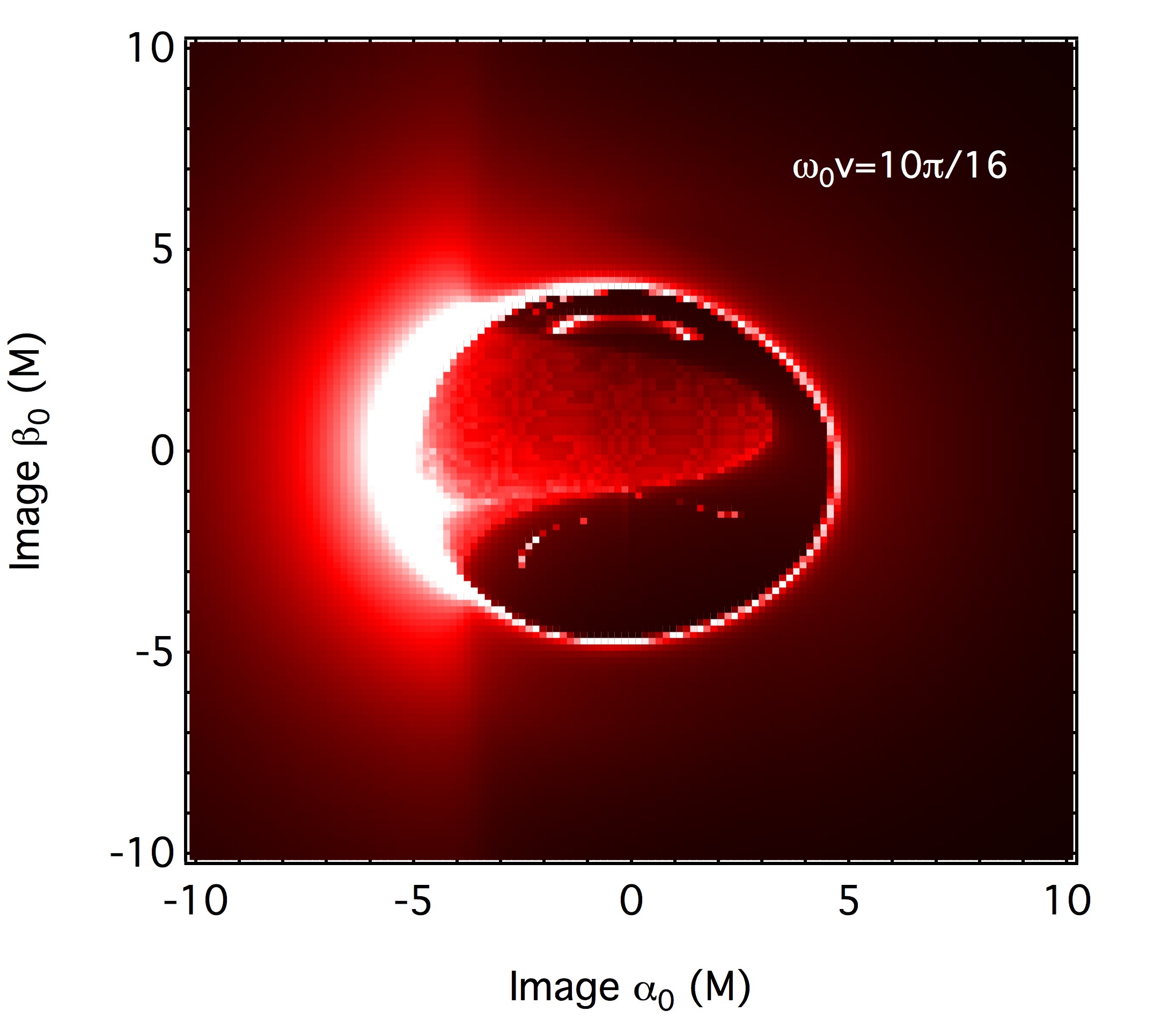}
    \includegraphics[width=0.5\textwidth]{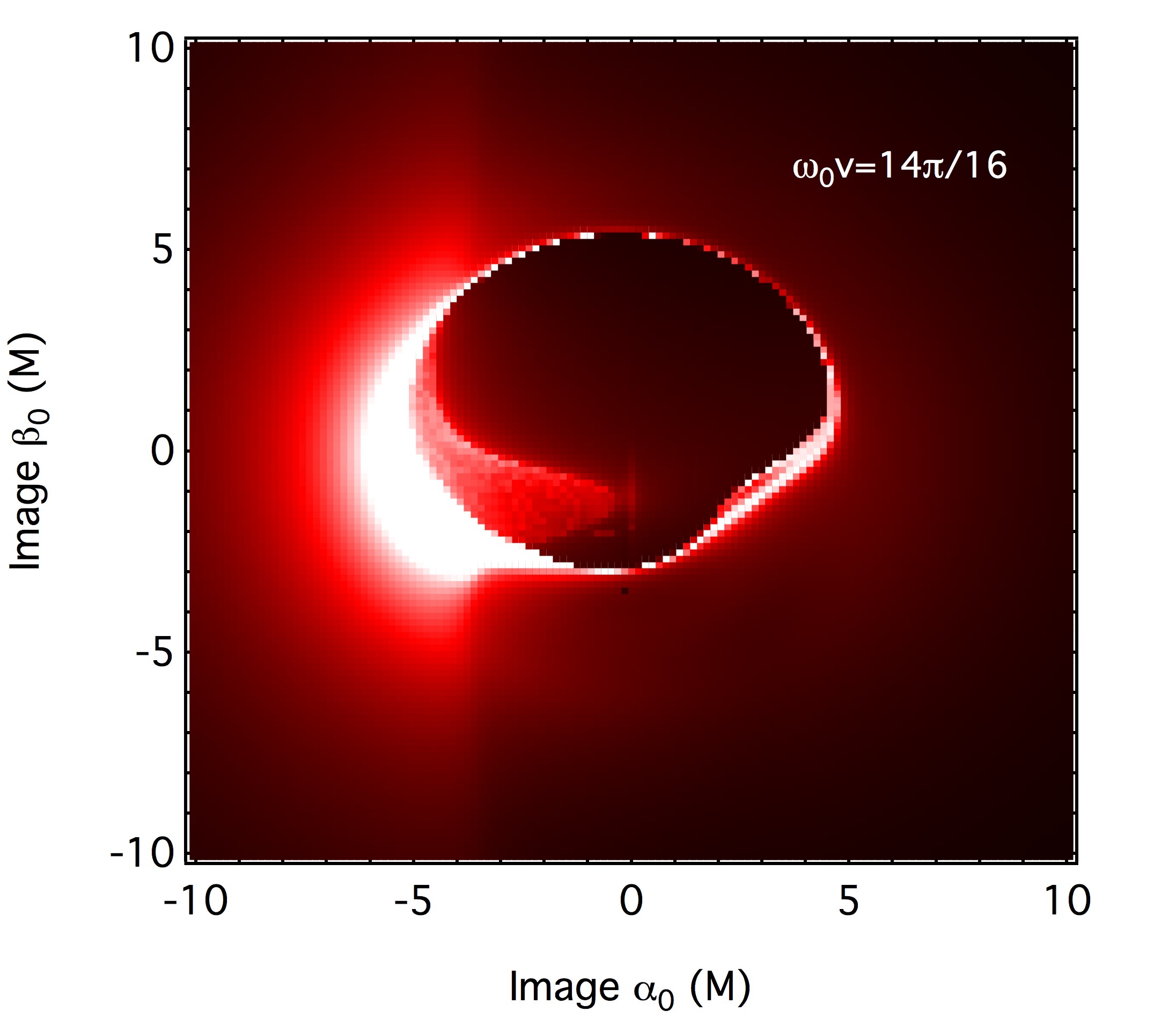} }
  \caption{Two snapshots of the time-dependent black-hole shadow
    calculated in Ref.~\cite{Giddings2016} for a spacetime that is
    characterized by quantum fluctuations at horizon scales.}
\label{fig:quantum}       
\end{figure*}

\noindent {\em Null Hypothesis Tests.---\/} Ref.~\cite{Psaltis2015}
proposed a null hypothesis test of the Kerr spacetime that is based on
the fact that the shadows of Kerr black holes have radii (in
gravitational units) that span only a very narrow range. The apparent
radius of a black hole shadow depends primarily on the ratio of the
mass $M$ to the distance $D_{\rm BH}$ of the black hole. For \sgra,
monitoring of the orbits of stars in the vicinity of the black hole
have constrained this ratio, which corresponds to the angular size in
the sky of one gravitational radius located at the distance of \sgra,
to
\begin{equation}
\frac{GM}{D_{\rm BH}c^2}=5.09\pm 0.17 \mu{\rm as}\;.
\end{equation}
Combining this measurement with the expected range of sizes of the
shadow for different black-hole spins and observer inclinations leads
to a prediction for the half angular size of the shadow of \sgra\ in
the sky of~\cite{Boehle2016}
\begin{equation}
  \delta \theta_{\rm Sgr A*}= 25.5\pm 0.9 \mu{\rm as}\;.
\end{equation}
This prediction has no free parameters. If EHT measurements of the
size of the black-hole shadow in \sgra\ find it inconsistent with this
prediction, then the null hypothesis (i.e., that \sgra\ is described
by the Kerr metric) will be falsified.

\bigskip

\noindent {\em Tests of the No-Hair Theorem and of non-Kerr
  metrics.---\/} Ref.~\cite{Johannsen2010a} proposed a test of the
no-hair theorem based on the fact that the shadow of a black hole is
nearly circular only if its spacetime obeys the particular relation
between the quadrupole moment and its spin that is dictated by the
no-hair theorem, i.e., if $q=-a^2$. If we allow for violation of the
no-hair theorem, i.e., allow for the quadrupole moment of the
spacetime to take arbitrary values, then the black-hole shadow becomes
asymmetric and its size can take significantly larger or smaller
values than what is given by equation~(\ref{eq:shadow_size}).  The
outlines of black-hole shadows have since been calculated for a large
number of metric that are either parametric extensions of the Kerr
metric or solutions to non-GR field
equations~\cite{sh21,sh20,sh19,sh18,sh17,sh16,sh15,sh14,sh13,sh12,sh11,sh10,sh9,sh8,sh7,sh6,sh5,sh4,sh3,sh2,sh1}
(see, e.g., Fig.~\ref{fig:nonKerrshadows}). The shadows for these
spacetimes become asymmetric even at small values of the spin because
the effect of the spacetime quadrupole and of frame dragging do not
nearly cancel each other. In other words, measuring the size and shape
of the black hole shadow and comparing the measurements to the values
predicted for the Kerr metric leads to a direct test of the no-hair
theorem.

\bigskip

\noindent {\em Tests of Quantum Structure.---\/} All the tests of
black-hole metrics discussed above search for deviations in the
structures of stationary spacetimes. It is plausible, however, that
the spacetimes of black holes appear to have classical dynamics
because of quantum fluctuations at horizon
scales~\cite{Giddings2017a}. Ref.~\cite{Giddings2016} explored the
impact of such fluctuations on the properties of the shadows cast by
black holes. They found that horizon-scale fluctuations of the
spacetime with even small amplitudes lead to order unity fluctuations
of the shape and size of the shadow (see Fig.~\ref{fig:quantum}). The
characteristic timescale of fluctuations would be $\sim 1$~hr for
\sgra\ and $\sim 60$~d for M87. Given that it takes several hours for
the EHT to generate a single image, it will require non-imaging
techniques to disentangle such signatures from the raw EHT data of
\sgra. On the other hand, such fluctuations will be readily visible in
the individual snapshots of M87 taken months or years apart.

\subsection{Implementation and Challenges}

The shape and size of the shadow of a black hole depends only on the
black-hole spacetime.  As a result, {\em tests of gravity that involve
  the properties of black-hole shadows are free from any astrophysical
  complications.\/} The presence of an accretion flow around the black
hole is necessary as a source of radiation on which the black hole can
cast its shadow. However, the intricacies of the thermodynamics of its
plasma, the structure of its turbulent magnetic fields, and its
emission properties, i.e., all of the issues that make predictions in
accretion physics difficult, do not affect the shadows of the black
holes.

The main complication introduced by the accretion flow is the fact
that it may be obscuring, partially or fully, the shadow. This would
have been especially important had the accretion flow been optically
thick~\cite{Luminet1979,Fukue2003}, which is not the case for the
primary targets \sgra\ and M87 (see discussion in \S2). However, even
for the radiatively inefficient flows of those black holes, the
left-right brightness asymmetry of the accretion flow (with respect to
its angular momentum axis) caused by relativistic Doppler effects (see
Figure~\ref{fig:grmhd}) and obscuration by the plasma intervening
between the observer and the event horizon (see, e.g.,
Ref.~\cite{Moscibrodzka2012,Medeiros2016a,Pu2016}) require special
care in the measurement of the shadow shape and size.

In principle, the properties of the black-hole shadow and, hence, of
the underlying spacetime, can be inferred indirectly by fitting models
of the accretion flow in different spacetimes to EHT
observations~\cite{Broderick2014}. This approach, however, depends
very strongly on the predictive power of accretion models and is,
therefore, susceptible to biases. To overcome this limitation, two
alternative approached have been suggested that focus entirely on
measuring characteristics of the shadow and not of the accretion flow.

Refs.~\cite{Johannsen2010a,Johannsen2016a} proposed measuring the
shape and size of the bright ring of light that surrounds the
black-hole shadow. This ring of light is the result of photon paths
that graze the various photon orbits and circle the black hole a very
large number of times before emerging towards a distant observer. The
total emissivity integrated along such photon paths becomes very large
and causes the narrow but bright ring of light surrounding the shadow
(see discussion in Ref.~\cite{Johannsen2010a}). In principle, the
narrow width of the bright ring allows for a precise measurement of
its shape and size~\cite{Johannsen2016a}. The applicability of this
approach, however, might be limited by the fact that the ring of light
will be hard to distinguish from the bright accretion flow image,
especially towards the side that is Doppler boosted towards the
observer and, hence, very bright.

A second characteristic of a black-hole shadow is the very abrupt
change in the image brightness across it. In other words, the outline
of the shadow of a black hole is the locus of points in the image with
the highest gradient in brightness. Ref.~\cite{Psaltis2015b} proposed
employing edge detection algorithms to identify the locations of
points on an image with the highest gradients and then applying a
pattern matching Hough/Radon transform in order to measure the
properties of the shadow. The benefits of this approach is that it
filters out all the flux that arises from the accretion flow and,
therefore, its associated complexities. However, for the case of
\sgra, this approach relies on an accurate mitigation algorithm for
the effects of interstellar scattering, which blurs the image and
smooths the sharp edge at the black-hole shadow.

The litmus test for any investigation on black-hole spacetimes that
relies on black-hole shadows is the verification that the same shadow
shapes and sizes are measured in repeated observations separated by
many dynamical timescales (hours for \sgra\ and months for
M87). Furthermore, because gravitational effects are achromatic
whereas plasma effects are not, the measurements of shadow shapes and
sizes will need to be consistent among observations at different
wavelengths and at different polarizations.  The EHT will be observing
its targets over multiple days during an observing season, over
multiple years, with polarization information, and at two wavelengths
(at least), i.e., 1.3~mm and 0.86~mm, offering many opportunities to
verify any measurements and to perform such consistency tests.

\begin{figure*}
  \centerline{ \includegraphics[width=0.5\textwidth]{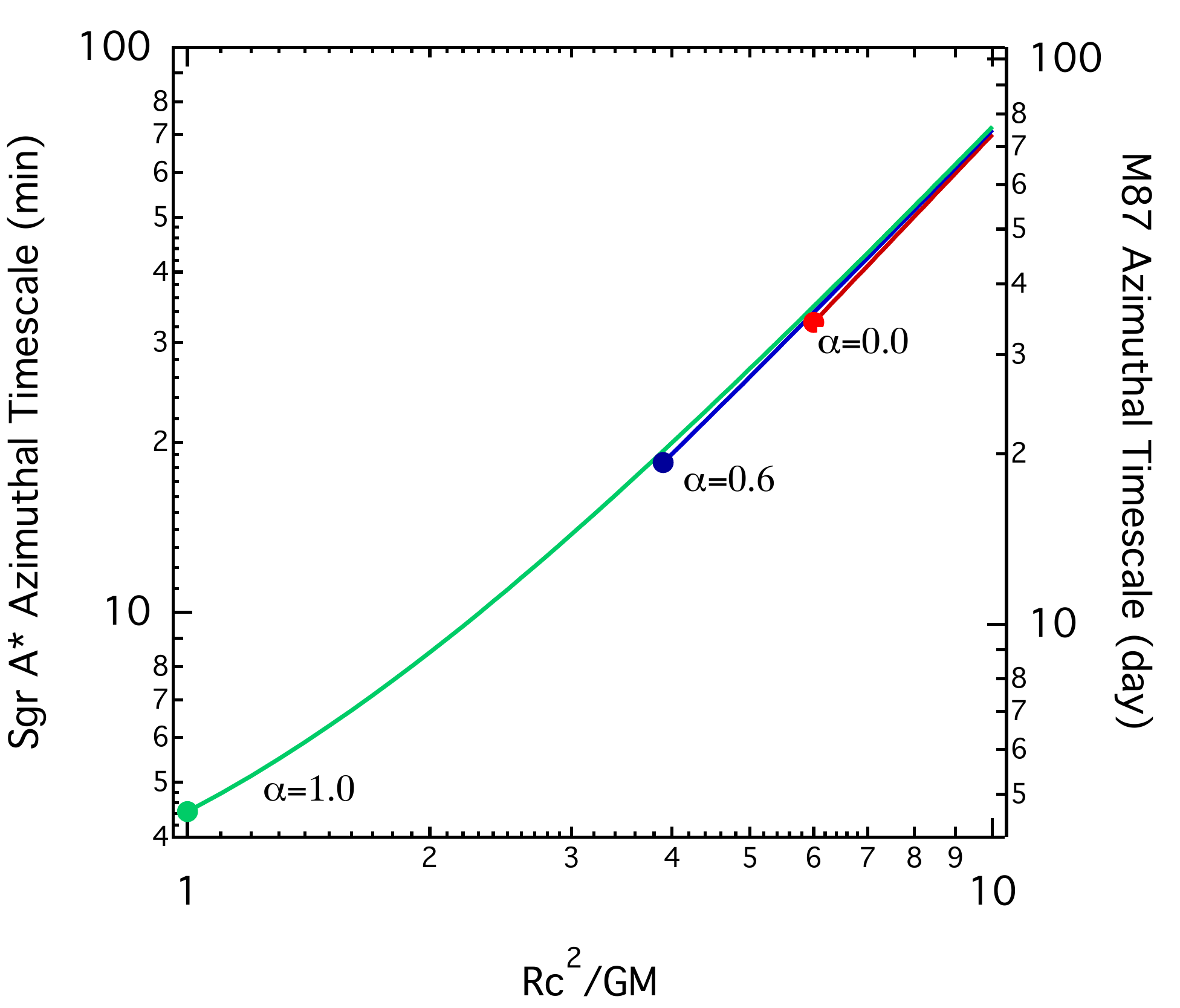}
    \includegraphics[width=0.5\textwidth]{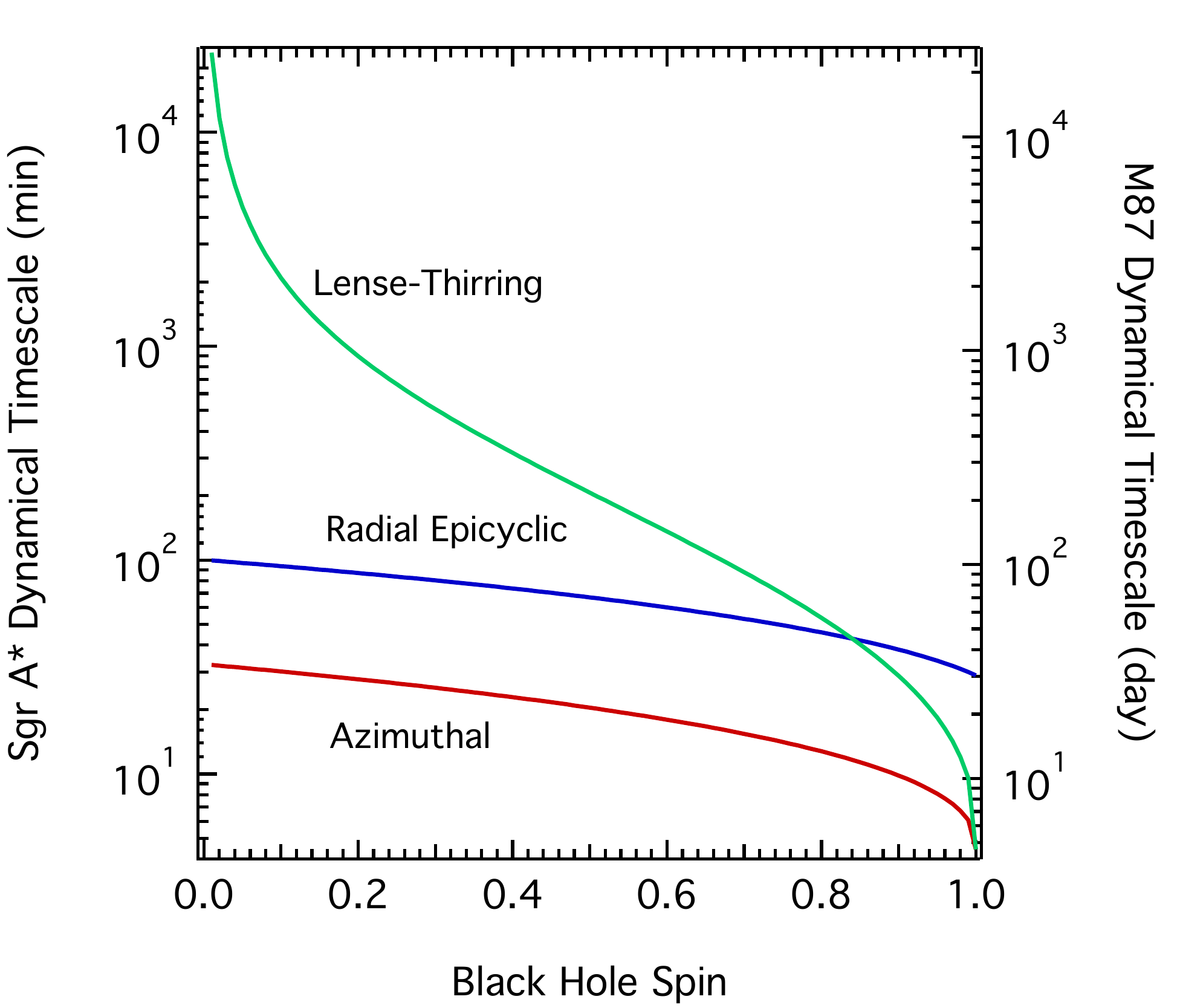} }
  \caption{{\em (Left)\/} The dynamical timescale for azimuthal
    (orbital) motions of test particles around a black hole, as a
    function of the location of the orbit, for three different values
    of the black-hole spin. The filled circles mark the location of
    the ISCO. {\em (Right)\/} The characteristic dynamical timescales
    for the azimuthal motion of a test particle at the ISCO, for the
    radial epicyclic motion at the location of the peak radial
    frequency, and for the Lense-Thirring precession at the location of
    the ISCO, as a function of the black-hole spin. The latter two
    timescales are expected to be comparable to the periods of
    $g-$ and $c-$modes excited in the inner accretion flows. In both
    panels, the left axes show the timescales in minutes for the mass
    of Sgr~A* and the right axes show the timescales in days for the
    mass of M87.}
\label{fig:freqs}       
\end{figure*}

\section{Tests with Timing Signatures}

The flux of radiation that emerges from accreting black holes has been
observed to be highly variable at all mass scales, from stellar-mass
black holes~\cite{Remillard2006} to supermassive black-holes, such as
\sgra~\cite{Genzel2003,Do2009,Neilsen2013}. This is expected given the
turbulent nature of the accretion flow and the caustic properties of
lensing in black-hole spacetimes~\cite{Rauch1994,Chan2015b}. For the
case of \sgra, observations in the infrared and in the millimeter have
shown that the variability is characterized by a broad band, red noise
spectrum with a potential turn-over at timescales longer than a few
hours~\cite{Meyer2009,Dexter2014a}. For M87, a similar turn-over
occurs at tens of days~\cite{Bower2015}.

In the case of stellar-mass black holes, observations can track
black-hole variability over millions of dynamical timescales, which
are as short as a few milliseconds. The surprising results of such
studies has been the discovery of quasi-periodic oscillations,
primarily in the X-ray flux, with very high quality
factors~\cite{Remillard2006} and regular and reproducible
properties~\cite{Psaltis1999}.  The origin of these oscillations is
not understood but their high coherences suggest that they are the
observational manifestations of linear~\cite{Wagoner1999,Kato2001} or
resonant~\cite{Abramowicz2003} oscillatory modes in the accretion
flows. This opens the possibility that the EHT images will show
compact coherent structures (e.g., the nodes of the oscillatory modes
that will appear as ``hot spots'') with fluxes or relative positions
on the images that oscillate in a quasi-periodic fashion for tens of
cycles. The expected frequencies of oscillations trace closely
dynamical frequencies in the black-hole spacetimes, offering the
possibility for additional gravity tests with EHT data.

\subsection{Properties of Timing Signatures}

We can define three dynamical frequencies at any given location in the
spacetime of a spinning black hole. For a test particle at a
Boyer-Lindquist radius $r$, the azimuthal frequency, which describes
the frequency of circular orbital motion as measured by an observer at
infinity, is given by~\cite{Bardeen1973}
\begin{equation}
  \Omega_{\phi}=\left(\frac{GM}{c^3}\right)^{-1}
  \frac{1}{(rc^2/GM)^{3/2}\pm a }\;,
\end{equation}
where, hereafter, the $\pm$ sign corresponds to prograde and
retrograde orbits.  The corresponding dynamical timescale for the two
primary targets becomes
\begin{eqnarray}
  \tau_{\phi}\equiv\frac{2\pi}{\Omega_\phi}
  &=&2.2\left(\frac{M_{\rm SgrA*}}{4.3\times 10^6 M_\odot}\right)
  \left[\left(\frac{rc^2}{GM_{\rm SgrA*}}\right)^{3/2}\pm a\right]~{\rm min}
  \nonumber\\
  &=&2.3\left(\frac{M_{\rm M87}}{6.5\times 10^9 M_\odot}\right)
  \left[\left(\frac{rc^2}{GM_{\rm M87}}\right)^{3/2}\pm a\right]~{\rm d}\;.
\end{eqnarray}
(Note that, because of the mass difference between \sgra\ and M87, one
minute for \sgra\ corresponds to one day for M87).

The azimuthal dynamical timescale is an increasing function of radius
(see also Fig.~\ref{fig:freqs}). Stable circular orbits exist only
outside the ISCO, the radius of which is given by~\cite{Bardeen1973}
\begin{equation}
  r_{\rm ISCO}=3+Z_2\mp\left[\left(1-Z_1\right)\left(3+Z_1+2Z_2\right)\right]^{1/2}\;,
  \label{eq:risco}
\end{equation}
where
\begin{eqnarray}
  Z_1&=&1+\left(1-a^2\right)^{1/3}\left[\left(1+a\right)^{1/3}
    +\left(1-a\right)^{1/3}\right]\\
  Z_2&=&\left(3a^2+Z1^2\right)^{1/2}\;.
\end{eqnarray}
As a result, the shortest azimuthal dynamical timescale that
corresponds to stable motions is equal to $\tau_\phi(r_{\rm ISCO}$)
and depends only on the black-hole spin. This is shown in the right
panel of Figure~\ref{fig:freqs}. For \sgra, it varies from $\sim
33$~min for zero spin to $\sim 4.4$~min for maximum spin and, for M87,
it varies between $\sim 30-4$~days. This is the fastest dynamical
timescale in an accretion flow and no significant variability is
expected to occur at faster timescales~\cite{Kluzniak1990}. Indeed,
this expectation has been verified in the case of stellar-mass black
holes~\cite{Sunyaev2000,Remillard2006}.

The radial epicyclic frequency, which describes the frequency of the
radial oscillations of a test particle in orbit, as measured by an
observer at infinity, is given by~\cite{Okazaki1987}
\begin{equation}
  \kappa=\Omega_\phi
  \left[1-6\frac{GM}{rc^2}-3a^2\left(\frac{GM}{rc^2}\right)^2
    \pm 8 a \left(\frac{GM}{rc^2}\right)^{3/2}\right]^{1/2}\;.
\end{equation}
For radii close to the black hole, the radial epicyclic frequency
becomes increasingly smaller than the azimuthal frequency. In fact,
the radial epicyclic frequency has a maximum at some characteristic
radius (equal to $8GM/c^2$ for a non-spinning black hole) and, by
definition, vanishes at the location of the ISCO. Linear gravity modes
(or g$-$modes) can be excited and trapped with frequencies comparable
to the maximum of the radial epicyclic frequency (see, e.g.,
Ref~\cite{Perez1997}). For a non-spinning black hole, the
characteristic dynamical timescale that corresponds to the maximum
radial epicyclic frequency (and, hence, to the fundamental g$-$mode) is
\begin{eqnarray}
  \tau_{r}\equiv\frac{2\pi}{\kappa}
  &=&100\left(\frac{M_{\rm SgrA*}}{4.3\times 10^6 M_\odot}\right)~{\rm min}
  \nonumber\\
  &=&105\left(\frac{M_{\rm M87}}{6.5\times 10^9 M_\odot}\right)~{\rm d}\;.
\end{eqnarray}
Figure~\ref{fig:freqs} shows that the radial timescale decreases with
increasing spin of the black hole and reduces to $\sim 30$~min for
\sgra\ and to $\sim 30$~d for M87.

Finally, the vertical epicyclic frequency, which describes the
frequency of vertical oscillations of a test particle in orbit, as
measured by an observer at infinity, is given by~\cite{Kato1990}
\begin{equation}
  \Omega_\perp=\Omega_\phi\left[1-4a\left(\frac{GM}{rc^2}\right)^{3/2}
    +3\alpha^2\left(\frac{GM}{rc^2}\right)^2\right]^{1/2}
   \;.
\end{equation}
Linear corrugation modes (or c$-$modes) can be excited at the
innermost regions of the accretion flow, with frequencies comparable
to the Lense-Thirring frequency (see, e.g., Ref.~\cite{Silbergleit2001})
\begin{equation}
\Omega_{\rm LT}\equiv\Omega_{\phi}-\Omega_\perp  
\end{equation}
that measures the precession frequency of the orbital plane of the
test particle. For a slowly spinning black hole, the characteristic
dynamical timescale that corresponds to the Lense-Thirring frequency
(and hence to the fundamental c$-$mode) at the ISCO is
\begin{eqnarray}
  \tau_{\rm LT}\equiv\frac{2\pi}{\Omega_{\rm LT}}
  &=&2400\left(\frac{a}{0.1}\right)^{-1}
  \left(\frac{M_{\rm SgrA*}}{4.3\times 10^6 M_\odot}\right)~{\rm min}
  \nonumber\\
  &=&2520\left(\frac{a}{0.1}\right)^{-1}
  \left(\frac{M_{\rm M87}}{6.5\times 10^9 M_\odot}\right)~{\rm d}\;.
\end{eqnarray}
Figure~\ref{fig:freqs} shows that the Lense-Thirring timescale
decreases rapidly with increasing spin of the black hole and actually
becomes shorter than the radial timesale for spins larger than $\sim
0.85$.

The timescales shown in Figure~\ref{fig:freqs} are comparable to the
expected periods of linear modes in the accretion flows that are
trapped in the frequency cavities dictated by the Kerr spacetime, with
small corrections due to hydrodynamic effects. However, it is plausible
that same modes at different locations and with different frequencies
may become resonant, reaching large observable
amplitudes~\cite{Abramowicz2003}. This appears to be the case for some
(but not all) pairs of simultaneous quasi-periodic oscillations
observed from stellar-mass black holes with frequencies in 3:2 ratios
(or similar ratios of small integers).

It is important to emphasize here that orbiting coherent structures
that lead to large-amplitude, quasi-periodic oscillations in the
images and brightness of accreting black holes have not been seen in
any of the numerical simulations discussed earlier. However, this is
most likely a shortcoming of the simulations. Nature somehow manages
to generate quasi-periodic oscillations in stellar-mass black holes
with structures that last for tens of cycles and modulate large
fractions (more than 10\%) of the total accretion luminosity.

\subsection{Proposed Tests}

Figure~\ref{fig:freqs} shows that the fundamental periods of different
accretion disk modes have different dependencies on black-hole
spin. As a result, identifying at least two of these modes and
measuring their frequencies leads to a measurement of the black-hole
spin~\cite{Wagoner2001,Abramowicz2001,Psaltis2004,Motta2014,Motta2014a}.
Similar to the case of helioseismology, each fundamental oscillatory
mode is accompanied by a larger spectrum of high-order modes, allowing
for the mode identification to be verified, if such a spectrum can be
detected..

\begin{figure*}
  \centerline{ \includegraphics[width=0.5\textwidth]{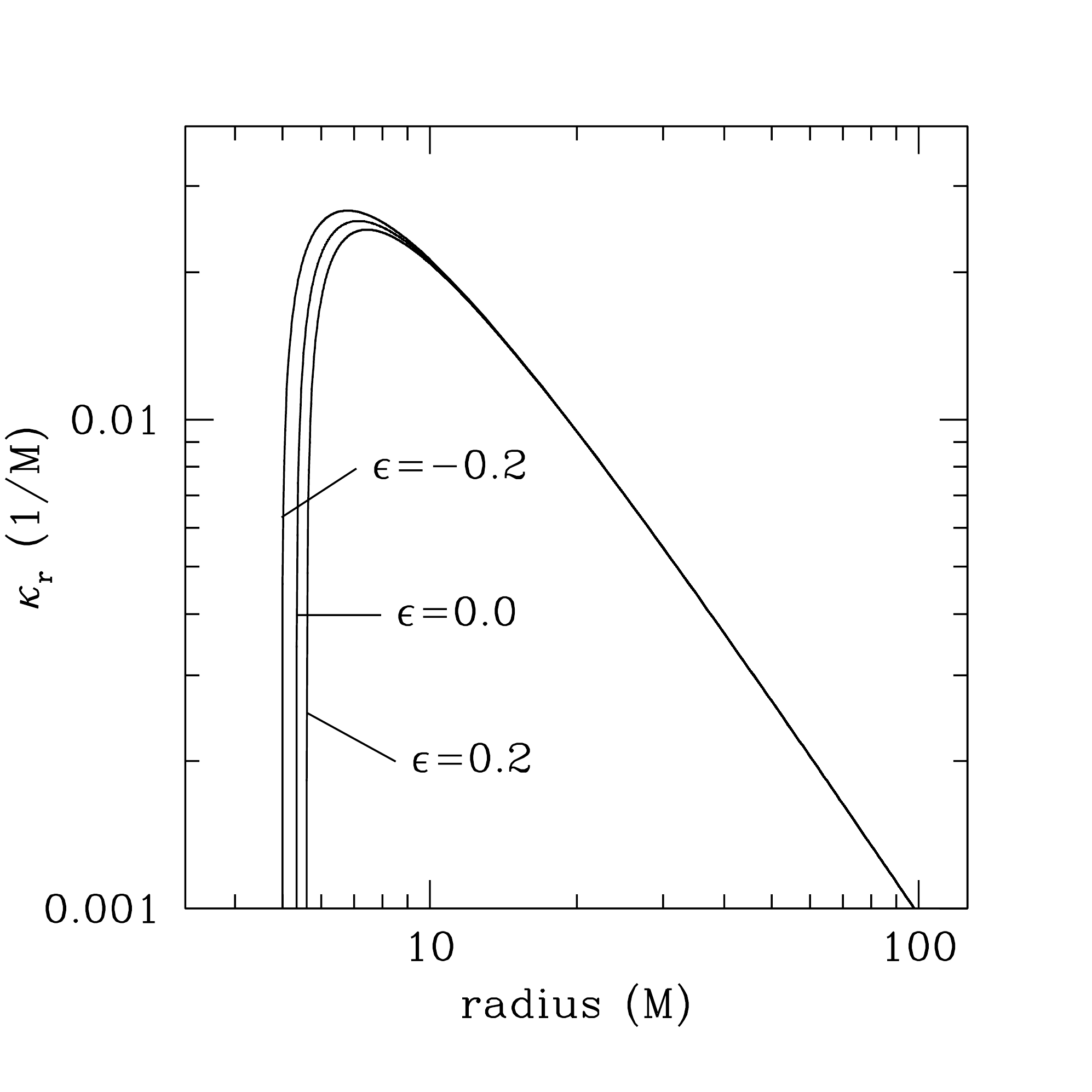}
    \includegraphics[width=0.5\textwidth]{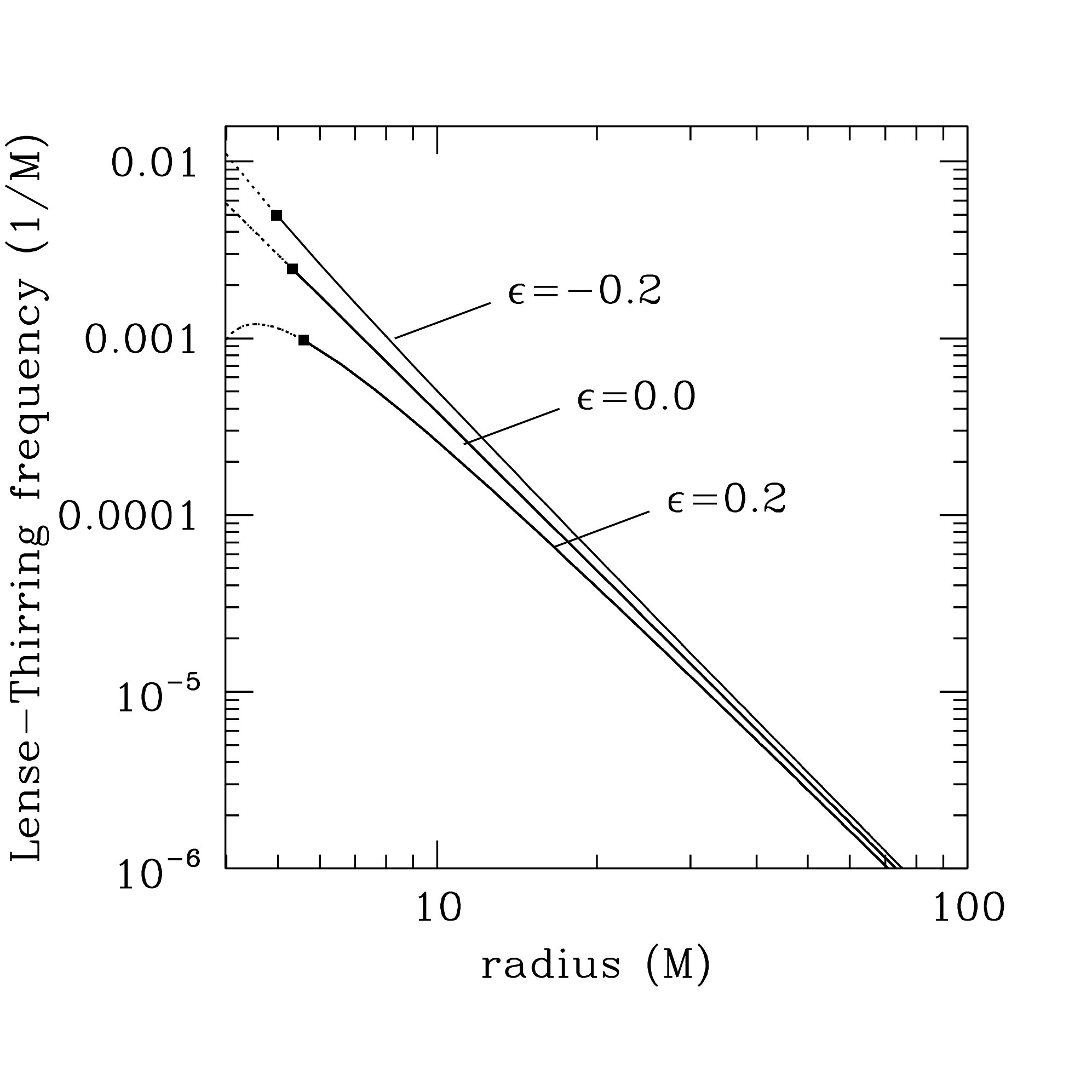} }
  \caption{The {\em (Left)\/} radial epicyclic and the {\em (Right)\/}
    Lense-Thirring frequency as a function of radius in the spacetime
    of an object that deviates from the Kerr metric. The parameter
    $\epsilon$ measures the degree to which the spacetime violates the
    no-hair theorem. (After Ref.~\cite{Johannsen2011}.)}
\label{fig:freq_epsilon}       
\end{figure*}

Ref.~\cite{Johannsen2011} proposed that identifying three modes around
a black-hole of known mass and measuring their frequencies can lead to
a test of the Kerr metric (see also~\cite{Bambi2012}). This test can be
cast, in principle, as a null-hypothesis test. In other words, two of
the frequencies can be used to measure the black-hole spin and this
information can then be used to predict the third frequency, with no
free parameters, and compare it to the observed value. Alternatively,
the same approach can be formulated as a test of the no-hair
theorem. To this end, Ref.~\cite{Johannsen2011} calculated the
characteristic dynamical frequencies in a quasi-Kerr spacetime with a
quadrupole moment that deviates by an amount $\epsilon=q+a^2$ from the
Kerr value (see Fig.~\ref{fig:freq_epsilon}). They found that the
azimuthal ($\Omega_\phi$) and the radial epicyclic frequencies
($\kappa$) show a very similar dependence on the deviation parameter
and can be used primarily to measure the black-hole spin. On the other
hand, the Lense-Thirring frequency, which corresponds to the frequency
of the c$-$mode shows an orthogonal dependence on the deviation
parameter and is optimal in measuring its value and verifying how
close to zero it is.

\subsection{Implementation and Challenges}

The EHT typically observes its two primary targets for a few hours
during a small number of nearly consecutive days and repeats the
observations 12 months later, when the weather conditions and the
elevation of the targets are optimal at all telescope
locations. Because the characteristic timescales of variability in M87
for, e.g., the azimuthal and radial modes range between 4 and 100 days
(see Figure~\ref{fig:freqs}), the cadence of observations is not
optimal for sampling several cycles of the expected variability.

The situation for \sgra\ is exactly the opposite. The corresponding
characteristic variability timescales range from 4 to 100 minutes (or
0.07 to 1.7 hours), which is nicely sampled by the cadence of
observations. However, these timescales are substantial shorter than
the several hours it takes for the Earth to rotate and the baselines
to cover a substantial fraction of the interferometric space to
generate an image. As a result, tests of gravity with timing
signatures in \sgra\ need to employ non-imaging techniques.

Ref.~\cite{Doeleman2009} proposed using the time variability of
closure phases along different baseline triangles to search for such
timing signatures (see also~\cite{Fraga2016}). A closure phase is the
sum of the complex visibility phases along three baselines that form a
closed triangle. Closure phases can be measured accurately with a fast
cadence and are independent of atmospheric delays and telescope gains,
both of which are hard to calibrate in mm VLBI. Moreover, closure
phases measure primarily the orientation, shape, and distance between
major bright regions in the image, making them optimal to search for
time periodicities in the image structure.

\begin{figure*}
  \centerline{ \includegraphics[width=0.55\textwidth,angle=-90]{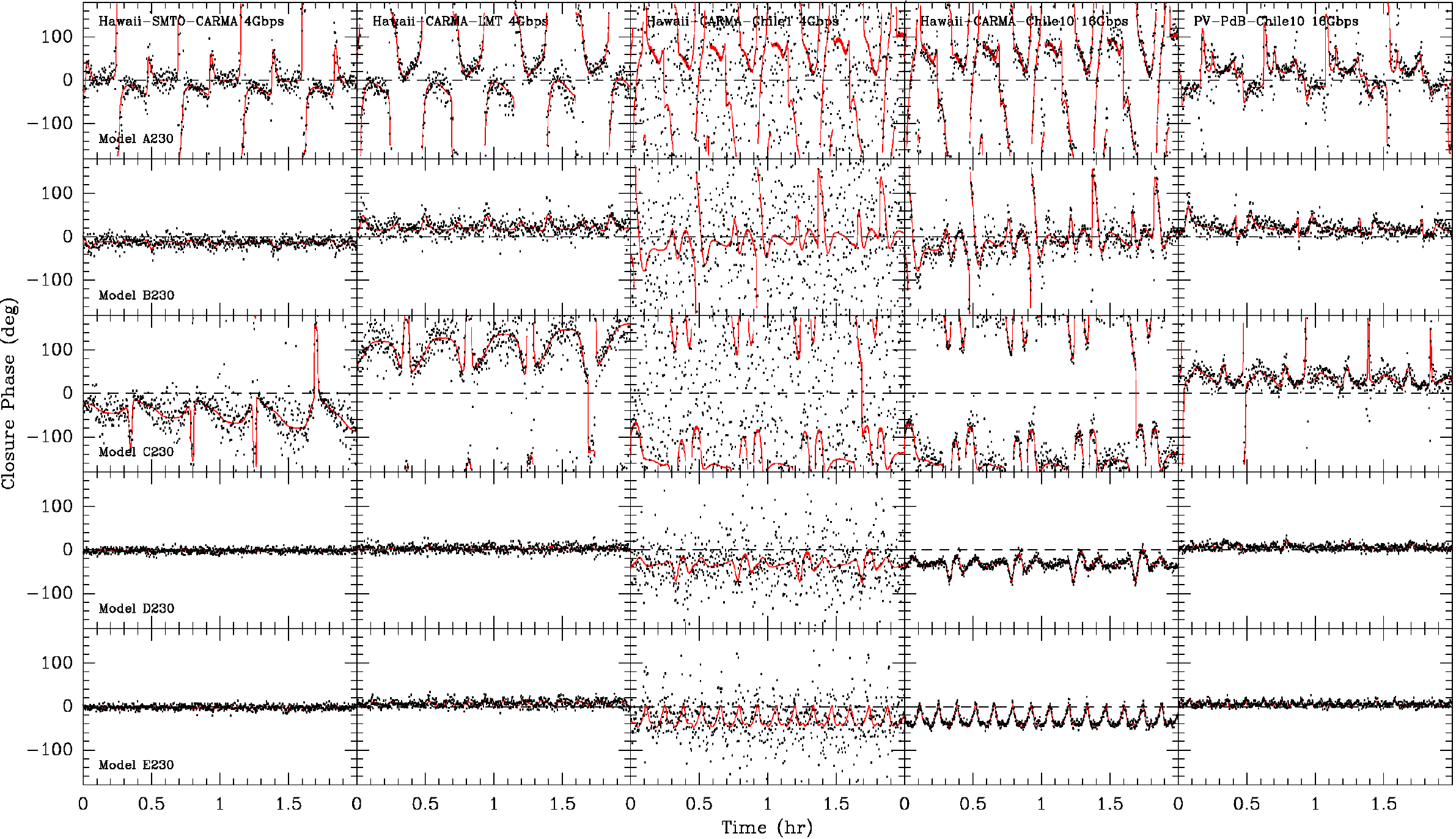}}
  \caption{Closure phases at different triangles of baselines {\em
      (columns)\/} and for different models {\em (rows)\/} of compact
    emission regions in orbit around \sgra. The detection of
    correlated, quasi-periodic oscillations in such closure phases
    with the Event Horizon Telescope may lead to measurements of the
    frequencies of oscillatory modes in the accretion flows and the
    properties of their spacetimes. (After Ref.~\cite{Doeleman2009}.)}
    \label{fig:closure}       
\end{figure*}

Figure~\ref{fig:closure} shows the effect of a number of example
models of orbiting ``hot spots'' around \sgra\ on the time evolution
of the closure phases along representative triangles of
baselines. Simple periodicity searches can easily detect quasi
periodicities in such signals as well as measure their frequencies and
coherence.

If such quasi-periodic signals are detected from \sgra\ with the EHT,
the main challenge of performing gravity tests with them will be in
identifying their physical origin, i.e., the oscillatory mode they
correspond to. As Figure~\ref{fig:freqs} shows, the detection of a
single period between 4-100~min can be attributed to different linear
modes or to different values of the black-hole spin. Even the
simultaneous detection of three different periods may not lead to a
conclusive result unless they are securely identified with particular
oscillatory modes or non-linear resonances.

The problem discussed above have severely hampered the ability to
perform similar tests of gravity with observations of quasi-periodic
oscillations from stellar-mass black holes (see discussion
in~\cite{Psaltis2004}). In the case of observations with the EHT, the
closure phases will provide not only measurements of the
characteristic oscillatory periods but also detailed information on
the relative sizes, shapes, and orientations of the structures that
cause them. Because each oscillatory mode corresponds to a very
particular structure in the accretion flow, it is possible that such
information will provide enough clues to identify each period with a
particular mode and lead to quantitative measurements of the
black-hole spin in \sgra\ and to tests of gravity.

\section{Combining Tests with the EHT, Stars, and Pulsars}

\begin{figure}
\centerline{\includegraphics[width=0.5\textwidth]{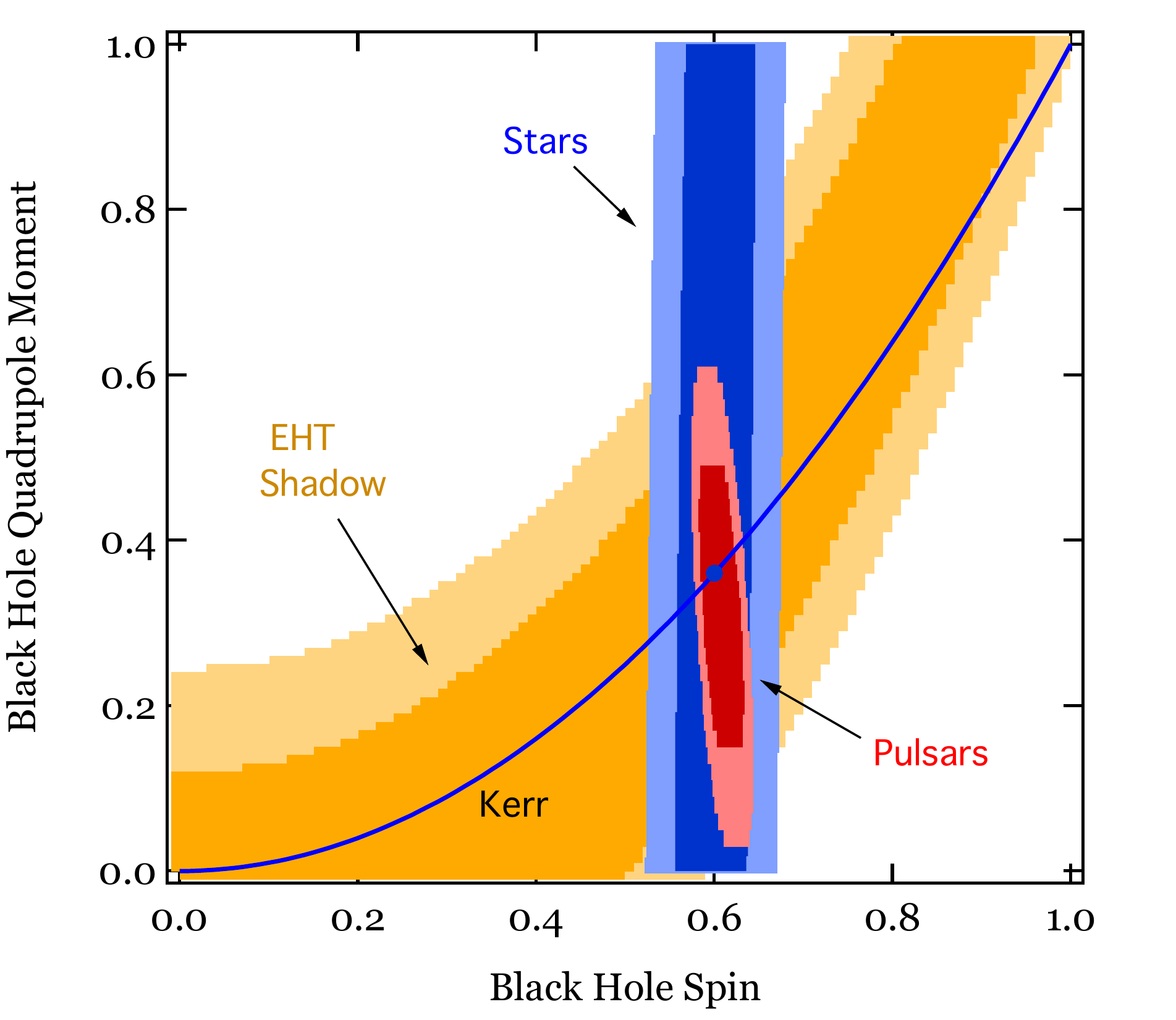}}
\caption{Combined constraints on the spin $a$ and spacetime quadrupole
  $q$ of \sgra\ based on hypothetical measurements of the shape and
  size of its shadow (orange), of the orbital precession of orbiting
  stars (blue), and of the timing properties of an orbiting pulsar
  (red).  If the black hole is described by the Kerr metric, the
  measurements should lie on the thin blue line for which $q=-a^2$.
  Each of these measurements faces different challenges and potential
  systematic effects. Statistical agreement between the three
  measurements will increase substantially their
  credibility~\cite{Psaltis2016}.}
\label{fig:pulsars}       
\end{figure}

As discussed in the previous sections, the EHT offers more than one
ways of testing gravity with its primary targets based on either
imaging or timing observations. In the case of \sgra, there are also
additional avenues of testing gravity using stars and pulsars that lie
on close elliptical orbits around the black hole~\cite{Psaltis2011}.

Ref.~\cite{Will2008} proposed a test of the no-hair theorem, i.e., an
independent measurement of the black-hole spin and quadrupole moment,
based on measuring the rate of precession of the periapsis and of the
plane of the orbits of stars. A number of studies have since explored
the requirements on the stellar orbits for such a measurement and the
potential complexities introduced by the presence of other stars and
gas in the vicinity of the black
hole~\cite{Merritt2010,Angelil2010,Sadeghian2011,Psaltis2012,Psaltis2013,Zhang2015,Psaltis2016,Yu2016,Waisberg2018}. There
is, indeed, a sweet spot of semi-major axes of $\sim 300-5000 GM/c^2$
for the orbits of stars that are optimal for measuring spacetime
parameters for \sgra. If such stars are discovered and monitored, they
will lead to measurements of the spin of the black hole with an
accuracy of $\sim 10$\% and a weak constraint on its quadrupole moment
(see Fig.~\ref{fig:pulsars})~\cite{Psaltis2016,Waisberg2018}.

Even though detecting the precession of a stellar orbit requires
continuous monitoring over multiple orbital periods and is susceptible
to various astrophysical complications, timing of even a single slow
pulsar in close orbit during a few periapsis passages will lead to an
accurate measurement of the spacetime moments of the black
hole~\cite{Liu2012,Angelil2014,Psaltis2016,Zhang2017}. This is true
because the time-of-arrival of the pulsar signal depends on the
properties of the spacetime along the line of sight from the current
location of the pulsar to the distant observer. As the relative
position of the pulsar, the black hole, and the spacetime evolve
during the periapsis passage, this allows mapping the spacetime and
measuring its properties, without requiring to wait for observing
actual orbital precession.

Figure~\ref{fig:pulsars} shows the combined constraints on the spin
$a$ and the quadrupole moment $q$ of \sgra\ based on a hypothetical
measurement of the shape and size of its shadow with the EHT, of the
orbital precession of orbiting stars, and of the timing properties of
an orbiting pulsar. As discussed in the previous sections, the shadow
measurement will constrain any possible deviations from the no-hair
relation $q=-a^2$; the precession of stellar orbits will measure
primarily the black-hole spin; and the timing of a pulsar will measure
the black-hole spin and its quadrupole moment. The three types of
measurements involve very different observations and techniques and
potentially suffer from uncorrelated biases. Moreover, they lead to
nearly orthogonal measurements of the spacetime moments. Performing
such measurements and generating results that are in statistical
agreement with each other will lead to highly credible tests of
gravity in the near horizon of an astrophysical, supermassive black
hole, with the Event Horizon Telescope.

\begin{acknowledgements}
  I thank my colleagues and students at the University of Arizona and
  especially T.\ Johannsen, J.\ Kim, L.\ Medeiros, C.\,K.\ Chan,
  F.\ \"Ozel and D.\ Marrone, who have contributed substantially to
  most of the work presented in this review. I also thank all the
  members of the EHT collaboration, who have made this incredible
  project a reality. Writing of this review has been supported by NSF
  PIRE grant 1743747.
\end{acknowledgements}

\bibliographystyle{spphys}       
\bibliography{eht}   

\end{document}